\documentclass[a4paper,11pt]{article}
\usepackage{graphicx}
\pdfoutput=1 

\usepackage{mathrsfs}
\usepackage[T1]{fontenc} 
\usepackage{bbold}
\usepackage{url}
\usepackage[english]{babel}
\usepackage[utf8]{inputenc}
\usepackage{amsmath}

\usepackage{amsfonts}
\usepackage{amssymb}
\usepackage{mathtools}
\usepackage{placeins}
\usepackage{float}
\usepackage{tabularx}
\usepackage{adjustbox}
\usepackage{caption}
\usepackage{subcaption}

\usepackage{jheppub} 

\title{\boldmath Tachyonic effects on K\"ahler moduli stabilized inflaton potential in type-IIB/F theory}

\author[]{Abhijit Let,}

\author{Buddhadeb Ghosh}

\affiliation[]{Centre of Advanced Studies, Department of Physics, The University of Burdwan,\\Burdwan 713 104, India}

\emailAdd{abhijitlet9692@gmail.com}
\emailAdd{bghosh@phys.buruniv.ac.in}

\abstract{We investigate the effects of inclusion of charged tachyonic open-string scalars in the perturbative and the non-perturbative K\"ahler moduli stabilizations in a geometry of three intersecting magnetized D7-brane stacks in type-IIB/F theory and also study the overall influence of this process on the inflaton potential, in a hybrid inflation scenario. We find that a tachyon lowers the minimum of the inflaton potential and assists to end the inflation. For simplicity, we have included one tachyon at a time in the present work and observe that this procedure preserves the features of slow-roll plateau of the potential. An interesting observation here is that the tachyonic part of the potential can be fine-tuned to get an almost zero minimum of the potential, thereby conforming to the small experimental value of the $\it{cosmological constant}$.}

\begin{document} 
\maketitle
\flushbottom

\section{Introduction}\label{sec1}

In recent years, cosmological inflation has been described through the K\"ahler moduli stabilizations in the framework of type-IIB/F theory with intersecting four-cycles involving magnetized $D7$-brane stacks, considering only the perturbative corrections \cite{Antoniadis:2018ngr,Antoniadis:2019doc,Antoniadis:2020stf} and both the perturbative and the non-perturbative corrections \cite{Let:2022fmu,Let:2023dtb}. Within the geometrical configuration consisting of three mutually orthogonal D7-brane stacks, there may exist open-string matter fields associated with these stacks. Some of the matter fields may become tachyonic under certain circumstances. The scenarios associated with these tachyonic fields have been discussed in \cite{Antoniadis:2021lhi,Antoniadis:2022ore}, at the backdrop of the perturbative logarithmic one-loop corrections.\par
Depending on the endpoint positions of open strings, which are attached to the D7-brane stacks and which acquire charges through U(1) gauge symmetry associated with the brane stacks, various types of matter fields such as twisted states ($D7^{(a)}-D7^{(b)}$), $a\neq b$ and untwisted states ($D7^{(a)}-D7^{(a)}$), where the indices $a$ and $b$ denote the brane stacks, may appear \cite{Antoniadis:2021lhi,Antoniadis:2022ore,Ibanez:2012zz,Lust:2004fi,Font:2004cx}. Due to the oscillator shifts of the charged open strings, owing to the magnetic fields of the D7-brane stacks, they get masses. The tachyonic effects of the twisted states can be eliminated by appropriate choice of the magnetic fields \cite{Antoniadis:2021lhi, Antoniadis:2022ore}. For the untwisted states, positive squared masses can come from either the brane separation (D-brane moduli) or the discrete Wilson lines. Some of the tachyonic states coming from the untwisted sector can be converted to matter fields (having positive squared masses) by appropriate choice of complex structure moduli, which come from the Wilson lines and the magnetic fluxes. \par
 The untwisted states, in conjunction with the brane separations, acquire positive squared masses when the volume of the internal space is greater than a critical value, which depends on the non-vanishing 
 VEV of the D-brane moduli. However, the brane separations give rise to the tachyonic fields when the volume of the internal space is less than some critical value. Here, the tachyonic fields behave as waterfall fields in a hybrid model \cite{Ahmed:2022dhc,Antoniadis:2021lhi,Gong:2022tfu} and assist to end the inflation by participating in the reheating mechanism \cite{Jain:2009ep,Copeland:2002ku, Rashidi:2017chf,Kofman:2001rb,Nautiyal:2018lyq}. Also, they provide a better cosmological constant by lowering the minimum value of inflaton potential. In this formalism, we have considered the D-brane moduli fields as stabilized by magnetic fluxes.

\par  
 In type-IIB string theory, compactified on Calabi-Yau three-fold, some of the massless scalar fields corresponding to the deformations of the compactified space are the dilaton field ($\Phi$), the K\"ahler moduli ($\tau_k$) and the complex structure moduli ($z_a$). Here, all the complex structure and the dilaton moduli are assumed to be stabilized by the three-form fluxes in the internal six-dimensional ($6d$) space at tree-level, whereas the K\"ahler moduli remain unstabilized \cite{Giddings:2001yu,Kachru:2003aw}. According to the KKLT model \cite{Kachru:2003aw}, all K\"ahler moduli are stabilized through the non-perturbative correction, by imposing the supersymmetric conditions and they acquire a scalar potential in the supersymmetric AdS vacua \cite{Kachru:2003aw, AbdusSalam:2020ywo,Carta:2019rhx}. The dS vacua are reached by an uplifting procedure with the addition of an anti-D3 brane. In contrast, in the LVS model \cite{Balasubramanian:2005zx} the K\"ahler moduli are stabilized by taking the combined effects of the perturbative and the non-perturbative corrections and an F-term scalar potential is realized with broken supersymmetry \cite{Balasubramanian:2005zx,Cicoli:2008va,Anguelova:2009ht,Cicoli:2011zz}. \par
 In the geometrical setup considered here, the D7-brane configurations themselves generate a D-term scalar potential for the K\"ahler moduli which can uplift the F-term potential from the $AdS$ to the $dS$ vacua \cite{Haack:2006cy,Antoniadis:2018hqy,Basiouris:2021sdf,Burgess:2003ic,Achucarro:2006zf,Kobayashi:2017zfd}. The logarithmic contribution which depends on the size of the two-dimensional space, transverse to the D7 brane, comes from the exchange of KK excitations between the localized gravity sources and the D7-branes in the large volume limit \cite{Antoniadis:2019rkh}. The magnetized D7-brane stacks wrapping the intersecting four cycles give rise to the non-perturbative contributions to the superpotential through gaugino condensations \cite{Basiouris:2020jgp,Basiouris:2021sdf}. The K\"ahler moduli, which appear in the superpotential, are fixed through supersymmetric minimizations and thus are treated as constants, appearing in the D-term potential \cite{Let:2022fmu,Let:2023dtb,Basiouris:2020jgp,Basiouris:2021sdf}. \par
  The main purpose of this paper is to examine the tachyonic effects of the open-string matter fields in the perturbative and non-perturbative moduli stabilization processes \cite{Let:2022fmu,Let:2023dtb}, insofar as the single-field cosmological inflaton potential is concerned. 
   The paper is organized in this way. In Section 2, we discuss the basic mechanism of closed string moduli stabilizations and give a brief overview of this aspect based on some previous works \cite{Antoniadis:2019rkh,Basiouris:2021sdf,Let:2023dtb}. In Section 3, we present a perspective of the open-string states and the appearance of the tachyonic fields in type-IIB string theory compactified on $T^6/(\mathbb{Z}_2\times\mathbb{Z}_2$) orbifold with three intersecting magnetized D7-brane stacks. Then, the inclusion of tachyonic fields in the computations of the F-term (subsection 3.1) and the D-term (subsection 3.2) potentials is analysed. In Section 4, we derive the inflaton potential including the tachyons, in terms of the canonically normalized inflaton field. Section 5 contains the results and the discussion for the present work. Finally, in section 6 we summarise the main results and write some concluding remarks.

\section{\boldmath Moduli stabilizations in type-IIB theory of closed strings and the scalar potentials}
In this section, we briefly review the moduli stabilization program (for closed strings) in a model containing three mutually orthogonal magnetized D7-brane stacks in the framework of type-IIB/F theory. For more details, one can see Refs.\cite{Basiouris:2020jgp,Basiouris:2021sdf,Let:2022fmu,Let:2023dtb}. The moduli-stabilized F-term potential, which is a combination of the K\"ahler potential and the superpotential, gives rise to the $AdS$ minima.  This potential is uplifted to the $dS$ minima through positive D-term contributions, which come from U(1) symmetries associated with the D7-brane stacks. Three K\"ahler moduli are related to the three intersecting magnetized D7-brane stacks. A zero-form potential (two-cycle axion, $C_0$) combines with the dilaton modulus ($\Phi$) to produce the usual axion-dilton modulus $S=C_0 + ie^{-\Phi}$. From the combination of the K\" ahler moduli and the four-cycle axions coming from the RR sector in type-IIB theory, we get complexified Kahler moduli \cite{Giddings:2001yu} ($\mathcal{T}_k$) as,
 \begin{equation}
     \mathcal{T}_k = b_k +i \tau_k ,                    \quad\quad k=1,2,3 \quad .
 \label{Eq:1}
 \end{equation}
 The internal volume, which is the Calabi-Yau three-fold ($CY_3$), can be expressed \cite{Antoniadis:2018hqy} in terms of the four-cycle volume moduli as, 
 \begin{equation}
     \mathcal{V} = \sqrt{\tau_1 \tau_2 \tau_3} =\sqrt{\frac{i}{8}\prod_{k=1}^3\left(\mathcal{T}_k -\Bar{\mathcal{T}}_k\right)} .
     \label{Eq:2.1}
 \end{equation}
  In this theory, a tree-level superpotential is induced by the three-form fluxes as proposed in \cite{Gukov:1999ya}. It depends on the complex-structure moduli ($z_a$) as, 
  \begin{equation}
    \mathcal{W}_0  = \int_{CY_3} G_3 \wedge \Omega {(z_a) } ,
    \label{Eq:W0}
 \end{equation}
 where, $ G_3 = F_3 - S H_3$ is a three-form flux which is a combination of field strengths $F_3 =dC_2, $ $ H_3 =dB_2$ and the axion-dilaton modulus (S). Here, $C_2$ and $B_2$ are a two-form potential and the Kalb-Ramond field, respectively. Also, the complex-structure moduli-dependent term $\Omega {(z_\alpha)}$ is a (3,0) holomorphic form. The K\"ahler moduli do not appear in the fluxed-induced superpotential of eq.(\ref{Eq:W0}).\par 
 
 In the effective field theory, another  important ingredient, which is known as K\"ahler potential, generates the metric of moduli spaces (complex-structure and K\"ahler moduli) of $ CY_3$ 
 and it depends logarithmically on various moduli fields \cite{Giddings:2001yu,Kachru:2003aw}. The tree level K\"ahler potential ($\mathcal{K}_0$), which is a no-scale type ($\kappa^2K^{\mathcal{T}_k\Bar{\mathcal{T}_k}}\partial_{\mathcal{T}_k} \mathcal{K}_0\partial_{\Bar{\mathcal{T}}_{k'}}\mathcal{K}_0=3$) for K\"ahler moduli ($\mathcal{T}_k$) is expressed as \cite{Antoniadis:2022ore,Basiouris:2021sdf,Let:2023dtb},
 \begin{equation}
     \kappa^2\mathcal{K}_0 =-2 \ln \left(\sqrt{\prod_{k=1}^3\frac{\left(\mathcal{T}_k -\Bar{\mathcal{T}}_k\right)}{(2i)^3}}\right) - \ln \big(-i(S-\Bar{S}\big)- \ln \Big(-i\int_{CY_3}\Omega\wedge\Bar{\Omega}\Big) ,
     \label{eq:2.6}
 \end{equation} where, $\kappa$ is the inverse of the reduced Planck mass.\par

     The general form of the F-term potential in $D=4,$ $ \mathcal{N}=1$ supergravity theory has the form \cite{Giddings:2001yu},
 \begin{align}
     V_F =e^{\kappa^2\mathcal{K}} \sum_{I,J} \Big(K^{I\Bar{J}}\mathcal{D}_I \mathcal{W}\mathcal{D}_{\Bar{J}}\mathcal{\Bar{W}}-3\kappa^2| \mathcal{W}|^2 \Big)
     \label{Eq:sf}
 \end{align}
 where $K^{I\Bar{J}}$ stands for the inverse metric of $ K_{I\Bar{J}} =\partial_I \partial_{\Bar{J}}\mathcal{K}$ and $\mathcal{D}\mathcal{W}=\partial \mathcal{W}+\kappa^2({\partial\mathcal{K}})\mathcal{W}$ is a covariant derivative. At tree-level, the F-term potential of eq.(\ref{Eq:sf}) can be written as
 \begin{equation}
     \begin{split}
         V_{\mathrm{F}}=e^{\kappa^2\mathcal{K}_0} \left(\sum_{a,b} K^{a,\Bar{b}}\mathcal{D}_a \mathcal{W}_0\mathcal{D}_{\Bar{b}}\mathcal{\Bar{W}}_0+\sum_{k=1}^3\kappa^2\big(\kappa^2K^{\mathcal{T}_k\Bar{\mathcal{T}_k}}\partial_{\mathcal{T}_k}\mathcal{K}_0\partial_{\Bar{\mathcal{T}_k}}\mathcal{K}_0-3\big)|\mathcal{W}_0|^2\right),
     \end{split}
     \label{Eq:TreF}
 \end{equation}
  where the indices $a$ and $b$ run over all complex-structure and dilaton moduli fields and $k$ stands for K\"ahler moduli.\par
 
  We focus on the K\"ahler moduli stabilization, assuming that all complex-structure and dilaton moduli have been stabilized through three-form fluxes ($G_3$) by imposing the supersymmetric condition on $\mathcal{W}_0$  and thus they are fixed with large masses \cite{Kachru:2003aw,Giddings:2001yu}. Due to the no-scale structure at the tree level, the scalar potential for all complex structures and dilaton moduli fields is obtained from eq. (\ref{Eq:TreF})  as,
 \begin{equation}
     V_{F}{\mathrm{(No-scale)}}=e^{\kappa^2\mathcal{K}_0} \sum_{a,b} K^{a,\Bar{b}}\mathcal{D}_a \mathcal{W}_0\mathcal{D}_{\Bar{b}}\mathcal{\Bar{W}}_0.
     \label{Eq:cd}
 \end{equation}
 This is identically zero due to the supersymmetric conditions $\it{viz}$, $\mathcal{D}_S \mathcal{W}_0=0$ and  $\mathcal{D}_{z_\alpha} \mathcal{W}_0 = 0$. At the tree level, no K\"ahler moduli fields appear in the scalar potential and thus they remain unstabilized. They are stabilized by the quantum corrections. It is described in \cite{Burgess:2005jx} that the flux-induced superpotential ($\mathcal{W}_0$) cannot receive any order of perturbative corrections due to non-renormalization theorem and thus only the non-perturbative contributions are effective in the process of stabilization.  
 
 \subsection{\boldmath Two non-perturbative contributions to the superpotential}
We have emphasized above that  $\mathcal{W}_0$ receives the non-perturbative corrections only, by which it becomes exponentially dependent on the K\"ahler moduli. There are several sources of non-perturbative corrections including the Euclidian D3-brane instantons wrapping four-cycles and gaugino condensations on D7-brane stacks  \cite{Kachru:2003aw,McAllister:2023vgy,Grana:2022nyp,Bena:2019mte}. In general, all K\"ahler moduli contribute to the superpotential and thus the superpotential becomes \cite{Balasubramanian:2004uy,Moritz:2017xto},
 \begin{align}
     \mathcal{W} = \mathcal{W}_0+\mathcal{W}_{np}=\mathcal{W}_0 +\sum_{k=1}^3 A_k e^{i a_k\mathcal{T}_k} 
 \label{Eq:2.9}
 \end{align}
 where, the index $k$ runs over all the K\"ahler moduli. The tree-level superpotential $\mathcal{W}_0$ is assumed to be a real constant \cite{Kachru:2003aw}. The coefficient $A_k$ depends on the complex structure moduli, which is also considered to be real. The value of the constant $a_k$ is $2\pi$ for instanton effects whereas for gaugino condensations it is $2\pi/N_k$, $N_k$ being the rank of the gauge group associated with the D7-brane stacks. Non-perturbative corrections involving both one K\"ahler modulus \cite{Basiouris:2020jgp,Let:2022fmu} as well as two K\"ahler moduli \cite{Let:2023dtb,Basiouris:2021sdf} have been considered, in the large volume limit. In the case of two non-perturbative contributions, which come from $\tau_1$ and $\tau_2$ moduli, the superpotential in eq.(\ref{Eq:2.9}) can be written as
 \begin{equation}
     \mathcal{W}= \mathcal{W}_0 + A e^{ia_1\mathcal{T}_1} +B e^{ia_2\mathcal{T}_2} . 
{\label{eq:non}}
\end{equation}
There are various coupling regions involving $\tau_1$ and $\tau_2$ as described in \cite{Basiouris:2021sdf,Let:2023dtb}. \par
In the present paper, we consider non-perturbative corrections of two of the three K\"ahler moduli. We must justify, here, the choice of non-perturbative corrections on either one or two moduli and not on all the three moduli. Primarily, large volume scenario (LVS) (see section 3.2 for its applications) and low-energy supergravity limit demand large transverse volumes of all the K\"ahler moduli, requiring $\tau_i\geq 1$. Making any of the moduli small would result in large curvature in K\"ahler cone \cite{Basiouris:2020jgp}, which will go against the LVS and the 
 supergravity limit. However, there is another mechanism, known as the `zero-mode lifting mechanism', proposed in \cite{Bianchi:2011qh}. In this mechanism, it is described that the worldvolume fluxes lift certain fermionic zero modes, which are superpartners of bosonic fields coming from geometrical deformations (known as moduli fields), and alter the effective action. As a consequence, a particular K\"ahler modulus is prevented from appearing in the non-perturbative superpotential. Here, we have made use of this mechanism in the case of the $\tau_3$ modulus (see \cite{Bianchi:2011qh} for details).  
 \subsection{K\"ahler potential with perturbative correction terms}
 Various types of perturbative corrections received by the K\"ahler potential break the no-scale structure of the tree-level K\"ahler potential $\mathcal{K}_0$ in eq.(\ref{eq:2.6}). In the 10d string theory, the perturbative corrections come from multi-graviton scattering amplitudes, which appear as the higher derivative terms in the low energy effective action \cite{Antoniadis:2019rkh,Green:1999pv,Basu:2006cs}. The $\alpha'^3$ (where, $\alpha'$ $=\sqrt{l_s}$ is the string Regge slope) contribution \cite{Becker:2002nn} causes a change in the volume term in eq.(\ref{eq:2.6}) by a constant shifting, $\xi =- \zeta(3)\chi/[4(2\pi)^3g_s^{3/2}]$, where $\chi$ is the Euler number of $ CY_3$. \par
 
 The present geometrical configuration, which is a combination of the intersecting D7 branes and O7 planes, gives rise to the volume-dependent logarithmic correction in K\"ahler potential \cite{Antoniadis:2019rkh,Basiouris:2020jgp,Basiouris:2021sdf,Let:2023dtb}. In the low energy expansion of type-IIB string theory  \cite{Green:1999pv,Green:2010wi}, the multi-graviton scattering amplitude upto one loop order induces $\mathcal{R}^4$ term in the 10d action \cite{Antoniadis:1997eg,Antoniadis:2019rkh,Antoniadis:2002tr,Kiritsis:1997em}, which, in turn,
    gives rise to the localized Einstein-Hilbert (EH) term $\mathcal{R}_{(4)}$ in four dimensions through dimensional reduction:  $ M_{10}=\mathcal{M}_4 \times \chi_{_6}$, where $\mathcal{M}_4$ is the 4d Minkowski spacetime and $\chi_{_6}$ is a 6-dimensional compactified manifold ($CY_3$). The coefficient of $\mathcal{R}_{(4)}$ is associated with the Euler number of $CY_3$, which is defined in terms of the 2-form Ricci curvature ($\mathcal{R}$)\cite{Antoniadis:2019rkh}, as 
 \begin{equation}
     \chi=\frac{3!}{(2\pi)^3} \int_{\chi_{_6}}\mathcal{R}\wedge  \mathcal{R} \wedge\mathcal{R} .
 \end{equation}
 The localized  EH terms exist in compactified 6-dimensional space associated with vertices for non-zero Euler numbers, which can emit massless gravitons and massive KK excitations of graviton-dilaton. The KK modes propagate on the transverse plane of the D7 branes. They are exchanged between some localized gravity sources on a torus (one-loop) and the  D7-branes/O7-planes, which give rise to the one-loop logarithmic contribution, dependent on the size of the compactified space. For the case of two massless gravitons and one KK excitation, the 4d supergravity action upto one-loop order is given by \cite{Antoniadis:2019rkh,Let:2022fmu,Let:2023dtb,Basiouris:2021sdf,Leontaris:2023obe},
 \begin{equation}
   \begin{split}
        S_{grav} &=\frac{1}{{(2\pi)}^7 \alpha'^4}\int_{\mathcal{M}_4 \times {\chi}_{_6}} e^{-2\phi}\mathcal{R}_{(10)}\\&+ \frac{\chi}{{(2\pi)}^4 \alpha'}\int_{\mathcal{M}_{4}}\left(2\zeta(3)e^{-2\phi}+4\zeta(2)\left(1 -\sum_k e^{2\phi}T_k \log\left(\frac{R^k_\perp}{\mathcal{L}}\right)\right) \right) {\mathcal{R}_{(4)}} ,
   \end{split}
 \label{Eq:mods}
 \end{equation}
 where, $T_k$ is the tension of the k-th D7 brane and for simplicity, we assume the same tension for all D7 branes ($T_k\equiv T=e^{-\phi}T_0$). Here, $R_\perp$ stands for the size of the two-dimensional space transverse to the D7 branes. The width of the localized source of gravity,  $\mathcal{L}\approx \frac{l_s}{\sqrt{N}}$ \cite{Antoniadis:2019rkh,Basiouris:2021sdf}, which implicitly depends on the Euler number of $ CY_3$, is deduced from graviton scattering involving massless gravitons and KK excitation \cite{Antoniadis:2019rkh,Antoniadis:2002tr,Leontaris:2023obe}. The integer N is related to the Euler number of $ CY_3$ ( $|\chi|\sim N$) and in the non-compact limit, it can be taken arbitrarily \cite{Basiouris:2021sdf}. The terms $\zeta(3)$ and $\zeta(2)$ come from the tree-level and the one-loop gravitons scattering amplitudes respectively\cite{Green:1999pv}.  In the non-compact limit ($R_\perp\rightarrow \infty$), the quantum corrected K\"ahler potential can be estimated \cite{Haack:2018ufg} from eq.(\ref{Eq:mods}) which takes the form   \cite{Antoniadis:2019rkh,Basiouris:2021sdf,Antoniadis:2018hqy},
 \begin{equation}
     \kappa^2\mathcal{K}(\mathcal{T}_k)=-2\ln{(\mathcal{V}+\xi+2\eta \ln{\mathcal{V}})} ,
 \label{Eq:k2}
 \end{equation}
 where, for smooth ${CY}$ (at the non-compact limit) case, the parameters $\xi$ and $\eta$ are given by \cite{Antoniadis:2019rkh,Antoniadis:2002tr,Antoniadis:2021lhi}: $\xi=-\frac{\zeta(3)}{4}\chi$, a positive constant (since $\chi<0$) and $\eta=-\frac{1}{2}g_sT_0\xi$, a negative constant (since $g_s,T_0,\xi>0$). The K\"ahler potential in eq.(\ref{Eq:k2}) breaks the no-scale structure and thus it induces non-zero F-term potential for K\"ahler moduli.

 \subsection{Scalar potential for three K\"ahler moduli}
The effective scalar potential for K\"ahler moduli is the sum of the F-term potential and the D-term potential. The F-term potential in eq.(\ref{Eq:sf}) can be written as,
\begin{equation}
    \begin{split}
        V_F=&e^{\kappa^2\mathcal{K}}\kappa^2\Big(\sum_{i,\Bar{j}=1}^3{K}^{\mathcal{T}_i{\Bar{\mathcal{T}_j}}}\Big(\kappa^{-2}\partial_{\mathcal{T}_i}\mathcal{W}\partial_{\Bar{\mathcal{T}_j}}\Bar{\mathcal{W}}+\mathcal{W}\partial_{\mathcal{T}_i}\mathcal{K}\partial_{\Bar{\mathcal{T}_j}}\mathcal{\Bar{W}}+\mathcal{\Bar{W}}\partial_{\mathcal{T}_i}\mathcal{W}\partial_{\Bar{\mathcal{T}_j}}\mathcal{K}\\&+\kappa^2\partial_{\mathcal{T}_i}\mathcal{K}\partial_{\Bar{\mathcal{T}_j}}\mathcal{K}|\mathcal{W}|^2\Big)-3|\mathcal{W}|^2\Big) 
\label{Eq:spil}
    \end{split}
\end{equation}
where $i,j=1,2,3$ run over all  K\"ahler moduli. In the large volume limit ($\eta,\xi\ll \mathcal{V}$), the K\"ahler potential in (\ref{Eq:k2}) reduces to,
 \begin{equation}
     \kappa^2\mathcal{K}(\mathcal{T}_k)=-2\ln{(\mathcal{V}+\mathcal{O}(\xi, \eta))}\approx -2\ln{\left(\sqrt{\frac{i}{8}(\mathcal{T}_1 -\Bar{\mathcal{T}}_1)(\mathcal{T}_2-\Bar{\mathcal{T}}_2)(\mathcal{T}_3-\Bar{\mathcal{T}_3)}}\right)}.
\label{Eq:lvs}
 \end{equation} The K\"ahler moduli $\tau_{1,2}$, which appear in superpotential, can now be stabilized supersymmetrically through the conditions:
\begin{equation}
\mathcal{D}_{\mathcal{T}_1}\mathcal{W}|^{\mathcal{T}_1 =i\tau_1}_{\mathcal{T}_2 =i\tau_2}= \mathcal{D}_{\mathcal{T}_2}\mathcal{W}|^{\mathcal{T}_1 =i\tau_1}_{\mathcal{T}_2 =i\tau_2}=0 \ .
\label{eq:super}
\end{equation}
  In the large volume limit (neglecting the term $\mathcal{O}(\xi,\eta)$), various coupling regions of $\tau_{1,2}$ moduli ($\beta a_2\tau_2e^{-a_2\tau_2}=a_1\tau_1e^{-a_1\tau_1} ; \beta=\frac{B}{A}$)\cite{Basiouris:2021sdf,Let:2023dtb} emerge from eq. (\ref{eq:super}). In the case of equal non-perturbative contributions coming from $\tau_{1,2}$ moduli, $\it{i.e.},$\cite{Let:2023dtb}  
  \begin{equation}
      a_1\tau_1\approx a_2\tau_2\geq \mathcal{O}(1) ,
      \label{Eq.a1a2}
  \end{equation}
we get,
 \begin{equation}
     -\left(a_1\tau_1+1\right)e^{-a_1\tau_1-1}=\frac{\gamma}{2e} \quad \Longrightarrow \omega=-(1+a_1\tau_1)=W_{0/-1}\Big(\frac{\gamma}{2e}\Big),
\label{Eq:a1}
\end{equation}
 where, the $W_{0/-1}$ denotes the `Lambert W-function'. The suffix `0' indicates  the lower branch (principle branch) and `-1' stands for the upper branch. The above equation implies the possible range of values of $\gamma$: $-2\leq\gamma<0$ (since, $\tau_{1,2}>0$).  \par
Eq.(\ref{Eq:a1}) leads to the following relations,
\begin{equation}
   \begin{split}
        \gamma=\frac{\mathcal{W}_0}{A}=2\omega e^{-a_1\tau_1}, \quad  \tau_1=-\frac{\left(1+\omega\right)}{a_1}.
        \label{Eq:sup}
   \end{split}
\end{equation}

The exact F-term potential \cite{Let:2023dtb} can be computed from eq. (\ref{Eq:spil}) by using eqs.(\ref{Eq:k2}) and (\ref{eq:non}). In the large volume limit and imposing the conditions in eq. (\ref{Eq:sup}), the F-term potential takes a simplified form, 
\begin{equation}
    V_F=\frac{\left(\epsilon\mathcal{W}_0\right)^2}{\kappa^4}\left(\frac{7\left(\xi+2\eta\ln\mathcal{V}-4\mathcal{V}\right)}{2\mathcal{V}^3}-\frac{17\eta\xi\ln\mathcal{V}}{\mathcal{V}^4}\right) ,
    \label{Eq.fk}
\end{equation}
where, $ \epsilon\left(=\frac{1+\omega}{\omega}\right)$ depends on the $\tau_1$ modulus. The F-term potential which gives  $AdS$ vacua,  cannot conform to the realistic universe. \par
In order to obtain the $dS$ vacua, a positive uplifting potential has to be added to the F-term potential. In type-IIB string theory, various uplifting mechanisms are applied depending on model \cite{McAllister:2023vgy,Dudas:2006vc,Bena:2018fqc,Dudas:2012wi,Dudas:2007nz,Dudas:2006gr}. For example, in the KKLT scenario, the introduction of an anti-D3-brane at the bottom of the Klebanov-Strassler (KS) throat gives rise to a positive D-term potential \cite{McAllister:2023vgy,Bena:2018fqc,Bena:2022ive,Dudas:2019pls}. The worldvolume fluxes of D7-brane stacks, which are associated with the U(1) gauge fields of the branes, generate a positive uplifting D-term potential (see \cite{Burgess:2003ic,Haack:2006cy,Achucarro:2006zf,Dudas:2006vc,Dudas:2005pr,Dudas:2005vv,Binetruy:1996uv,Freedman:2012zz} for details). In $N=1$ and D=4 supergravity theory, the general expression of the D-term  can be written as \cite{Haack:2006cy,Burgess:2003ic,Achucarro:2006zf,Dudas:2006vc}, 
\begin{equation}
    \begin{split}
        V_D=\sum_{k=1}^3\frac{g_k^2}{2}\left(iQ_k\partial_{\mathcal{T}_k}\mathcal{K} +\sum_j q_j|\left<\Phi_j\right>|^2\right)^2 ,
    \label{Eq.Do}
    \end{split}
\end{equation}
where $g_k$'s ($g_k^{-2}= \tau_k +$ flux and curvature part containing dilaton) stands for gauge coupling of D7 branes and $Q_k$ is the charge of four-cycle volume modulus $\tau_k$ acquired from the U(1) gauge symmetry. Here, the $q_j$ are the charges of the matter fields $\Phi_j$, which come from an open string attached to the D3 and D7 branes (D3-D7 strings, see \cite{Burgess:2003ic,Dudas:2006vc,Haack:2006cy,Achucarro:2006zf} for details). In general, The VEVs of the matter fields $\left<\Phi_j\right>$ are non-vanishing, but it may be possible to assume them to be zero in the cases of either the separation of D3-D7 branes being very large, which implies the enormous masses of corresponding open string states, or considering only D7 branes with fluxes but no D3 branes \cite{Burgess:2003ic}. However, this scenario is model-dependent. In our work, which is based on the model \cite{Antoniadis:2018hqy,Basiouris:2020jgp,Basiouris:2021sdf,Let:2023dtb} containing only D7-branes stacks (at least three intersecting D7-brane stacks \cite{Antoniadis:2022ore}) with fluxes, the VEVs of the matter fields are assumed to be zero. Then, the D-term potential in eq. (\ref{Eq.Do}) takes the form \cite{Antoniadis:2018hqy,Basiouris:2021sdf,Let:2023dtb} as,
\begin{equation}
    V_D=\sum_{k=1}^3\frac{g_k^2}{2}\left(iQ_k\partial_{\mathcal{T}_k}\mathcal{K}\right)^2\approx \sum_{k=1}^3\frac{d_k}{\kappa^4\tau_k^3} ,
    \label{Eq.D}
\end{equation}
  where, $d_k$, which are proportional to $Q_k^2 (>0)$, are positive constants.\par
  The effective potential ($V_{\mathrm{eff}}=V_{\mathrm{
F}}+V_{\mathrm{D}}$) is obtained from eqs. (\ref{Eq.fk}) and (\ref{Eq.D}) as, 
\begin{equation}
   \begin{split}
        V_{\mathrm{eff}}=\frac{\left(\epsilon\mathcal{W}_0\right)^2}{\kappa^4}\left(\frac{7\left(\xi+2\eta\ln\mathcal{V}\right)-4\mathcal{V}}{2\mathcal{V}^3}-\frac{17\eta\xi\ln\mathcal{V}}{\mathcal{V}^4}\right)+\frac{d_1}{\kappa^4\tau_1^3}+\frac{d_3}{\kappa^4\tau_3^3}+\frac{d_2\tau_1^3\tau_3^3}{\kappa^4\mathcal{V}^6}.
    \label{Eq:effmb}
    \end{split}
\end{equation}
  where, the $\tau_2$ modulus has been replaced by $\tau_{1,3}$ and $\mathcal{V}$ through eq. (\ref{Eq:2.1}). The moduli $\tau_{1,2}$ are stabilized supersymmetrically through the eq.(\ref{eq:super}) and thus, they are treated as  constants (eq.(\ref{Eq:a1}))\cite{Basiouris:2020jgp} (see also \cite{Randall:2019ent,Brummer:2005sh}). There is only one variable (either $\tau_3$ or $\mathcal{V}$) present in effective potential, eq.(\ref{Eq:effmb}). The effective potential of eq. (\ref{Eq:effmb}) can be minimized with respect to either the $\tau_3$ modulus or the $\mathcal{V}$ \footnote{In the present model of perturbative and non-perterbative corrections, the detailed procedures of K\"ahler moduli stabilization have been described in \cite{Basiouris:2020jgp,Basiouris:2021sdf}.}. Here, we do the minimization of the effective potential  with respect to the modulus $\tau_3$ and get \cite{Let:2023dtb}, 
  \begin{equation}
    \begin{split}
        \frac{dV_{\mathrm{eff}}}{d\tau_3}=0\quad\Longrightarrow\quad(\tau_3)_{\mathrm{min}}=\left(\frac{d_3}{d_2}\right)^{1/6}\frac{\mathcal{V}}{\tau_1^{1/2}}.
    \end{split} \label{Eq:ta3m} 
  \end{equation} Using eq. (\ref{Eq:ta3m}) in eq. (\ref{Eq:effmb}), we get, 
\begin{equation}
    V_{\mathrm{{eff}}}\left(\mathcal{V}\right)=\frac{\left(\epsilon\mathcal{W}_0\right)^2}{\kappa^4}\left(\frac{7\left(\xi+2\eta\ln\mathcal{V}\right)-4\mathcal{V}+p'}{2\mathcal{V}^3}-\frac{17\eta\xi\ln\mathcal{V}}{\mathcal{V}^4}\right)+\frac{V_{\mathrm{up}}}{\kappa^4}
    \label{Eq:effm}
\end{equation}
where, $p'=\frac{2d}{\left(\epsilon\mathcal{W}_0\right)^2}$, $V_{\mathrm{up}}=\frac{d_1}{\tau_1^3}$, $d=2\tau_1^{3/2}\sqrt{d_2d_3}$ and  are the constants. The inflaton field, $\phi$, can be obtained by canonical normalization of the K\"ahler modulus $\tau_3$ and thus we get,
\begin{equation}
    \phi=\frac{1}{\sqrt{2}}\ln{\left(\tau_1\tau_2\tau_3\right)}=\sqrt{2}\ln{\mathcal{V}}\quad\Longrightarrow\quad \mathcal{V}=e^{\frac{1}{\sqrt{2}}\phi}.
    \label{Eq.canonical}
\end{equation}
Finally, the inflaton potential is written from eq.(\ref{Eq:effm})  as,
\begin{equation}
    V\left(\phi\right)=\frac{\left(\epsilon\mathcal{W}_0\right)^2}{2\kappa^4}e^{-\frac{3}{\sqrt{2}}\phi}\left(p -4e^{\frac{1}{\sqrt{2}}\phi}+\left(7-17\xi e^{-\frac{1}{\sqrt{2}}\phi}\right)\sqrt{2}\eta\phi\right)+\frac{V_{\mathrm{up}}}{\kappa^4},
    \label{Eq.wotach}
\end{equation}
where, 
\begin{equation}
    \begin{split}
        p=p'+7\xi=\frac{4\tau_1^{3/2}\sqrt{d_2d_3}}{(\epsilon\mathcal{W}_0)^2}+7\xi \quad \mathrm{and}\quad V_{\mathrm{up}}=\frac{d_1}{\tau_1^3}.
    \end{split}
    \label{Eq.pup}
\end{equation}
 Here, $p$ and $V_{\mathrm{up}}$ are volume ($\mathcal{V}$)-independent constants. The uplifting term,  $V_{\mathrm{up}}$,  comes from the D-term potential by supersymmetrically fixing of $\tau_{1,2}$ and it has an important role to play in obtaining the slow-roll inflation. In our previous work \cite{Let:2023dtb}, the cosmological parameters were calculated with an appropriate choice of the values of $\mathcal{W}_0,\eta,\xi,\epsilon,d,d_1,\tau_1,a$. \par
 As we shall see in the following section, if we include the charged open string sector associated with the D7 brane stacks into our scheme, they will become tachyonic under the influence of a number of factors such as the magnetic fields of the branes, the brane separation and the internal volume. For a particular choice of the first two factors, the tachyonic effects come through a phase transition at a critical value of the internal volume.

\section{\boldmath Matter fields and scalar potential}
\subsection{Conditions for obtaining the tachyonic matter fields}
In the previous section, we have reviewed our earlier works on stabilization of three K\"ahler moduli, all of them perturbatively upto one-loop order and some of them non-perturbatively (instanton or gaugino condensation effects). This was to prepare the background on which the main works in the present paper, $\it{viz}$, consideration of open string matter fields associated with the D7-brane stacks in the moduli stabilization process, over and above the non-perturbative and the perturbative corrections, will be described. Since the open string scenario in the present geometrical setup has been considered somewhat extensively in recent literature \cite{Antoniadis:2022ore,Antoniadis:2021lhi}, we shall be brief, here, while describing the formalism.\par
In the case of $T^6/\mathbb{Z}_2\times\mathbb{Z}_2$ orbifold compactification \cite{Antoniadis:2021lhi,Antoniadis:2022ore,Ibanez:2012zz,Font:2004cx}, $T^6$ is actually an orthogonal combination of three two-cycle  tori:   $T^6=T_1^2\times T_2^2\times T_3^2$. The $D7^{(a)}$-brane stack lies on the plane of $T_a^2$ torus and is wrapped by $T_j^2, T_k^2$ ($a\neq j\neq k$) tori.   The modes of the open string charged oscillators connected with the $D7^{(a)}$-brane stack are shifted by worldvolume magnetic field ($H_k^{(a)}$) along the $D7_k$ plane, so that the stack moves in the worldvolume spanning the $D7_j$  plane. The shift is given by \cite{Antoniadis:2021lhi,Ibanez:2012zz,Bachas:1995ik,Antoniadis:2022ore,Font:2004cx},
 \begin{equation}
     \zeta_k^{(a)}=\frac{1}{\pi}Arctan\left(2\pi\alpha' q^{(a)}H_k^{(a)}\right)
     \label{Eq.os}
 \end{equation}
where, $q^{(a)}=\pm 1$ are the charges at the open string endpoints which are acquired by U(1) gauge symmetry. The magnetic field breaks supersymmetry and as a consequence, the oscillators get masses\cite{Bachas:1995ik}. The mass of open string states can be computed through the mass-shift formula \cite{Antoniadis:2021lhi,Bachas:1995ik,Angelantonj:2000hi} given by, 
\begin{equation}
    2\alpha' m^2=\sum_i \left[\left(2n_i+1\right) \left|\zeta_i^{(L)}+\zeta_i^{(R)}\right|+2\Sigma_i\left(\zeta_i^{(L)}+\zeta_i^{(R)}\right)\right]
    \label{Eq.ms}
\end{equation}
where, the subscripts $L,R$ stand for open string left and right endpoints which are attached with the D7-brane stacks and they have to be replaced by the corresponding branes in the oscillator shift in eq.(\ref{Eq.os}). $\Sigma_i$ are the internal helicities \cite{Antoniadis:2022ore,Bachas:1995ik}.\par
Based on the geometry of the three intersecting D7-brane stacks \cite{Antoniadis:2021lhi,Antoniadis:2022ore}, different kinds of open-string states appear depending on the positions of their endpoints.  Some are the untwisted doubly charged states ($D7^{(a)}-D7^{(a)}$) ($\it{i.e.},$ both endpoints have the same charge) and some are the twisted charged states ( $D7^{(a)}-D7^{(b)}$ with a,b=1,2,3 and $a\neq b$). The masses of the states ($D7^{(a)}-D7^{(a)}$) ($\it{i.e.},$ lowest-lying untwisted states) can be deduced \cite{Antoniadis:2021lhi,Antoniadis:2022ore,Font:2004cx} from the mass shift formula in eq.(\ref{Eq.ms}) as, 
\begin{equation}
    \begin{split}
        \alpha'(m^{(a)})^2=-2|\zeta_i^{(a)}|,\quad \mathrm{where}, \quad i\neq a,
        \label{Eq.tach}
    \end{split}
\end{equation}  
whereas, the masses of the lowest-lying charged twisted states are 
\begin{equation}
    \begin{split}
     &\alpha'(m^{(12)})^2=|\zeta_3^{(2)}|-|\zeta_2^{(1)}| \\& \alpha'(m^{(13)})^2=|\zeta_1^{(3)}|-|\zeta_2^{(1)}|\\&\alpha'(m^{(23)})^2=|\zeta_1^{(3)}|-|\zeta_3^{(2)}|.
    \end{split}
\end{equation}
The masses of the twisted states can be made zero by choosing $|\zeta_2^{(1)}|=|\zeta_3^{(2)}|=|\zeta_1^{(3)}|$. The magnetic fields $H_i^{(a)}$ can follow the Dirac quantization condition \cite{Angelantonj:2000hi}: $2\pi v_iH_i^{(a)}w_i^{(a)}=n_i^{(a)}$, where $n_i^{(a)}$ is the flux number and  $w_i^{(a)}$ the wrapping number of the $D7^{(a)}$-brane stacks on the $i$-th internal plane ($T_i^2$). For simplicity and to maintain the U(1) gauge symmetry of the separated D7-brane stacks, the corresponding wrapping numbers will be considered to have the value 1 \cite{Antoniadis:2022ore}. Here, $v_i$ is a two-cycle modulus, related to the internal volume as $\mathcal{V}=\alpha'^{-3}v_1v_2v_3$. In the small magnetic field approximation  ($\it{i.e.},$ in the large volume limit) the eq.(\ref{Eq.os}) becomes,
\begin{equation}
    \begin{split}
      \zeta_i^{(a)}= \frac{1}{\pi} Arctan\left(\frac{\alpha'\kappa_i^{(a)}}{v_i}\right)\approx \frac{\alpha'\kappa_i^{(a)}}{\pi v_i} \quad \mathrm{where}, \quad \kappa_i^{(a)}=\frac{n_i^{(a)}}{w_i^{(a)}}.
    \end{split}
\end{equation}
\par
The tachyonic states in eq.(\ref{Eq.tach}) can be uplifted to positive-squared-mass matter fields by the Wilson lines or by introducing distances of separations between the branes and their images in the transverse direction \cite{Antoniadis:2022ore,Antoniadis:2021lhi,Ibanez:2012zz}. 
First, we consider the case of the Wilson lines. For example, for discrete Wilson lines along the third and the first tori for the $D7^{(1)}$ and $D7^{(2)}$ stacks respectively, the masses of ($D7^{(1)}-D7^{(1)}$) and ($D7^{(2)}-D7^{(2)}$) states  are modified as \cite{Antoniadis:2021lhi,Antoniadis:2022ore},
\begin{equation}
    \begin{split}
        &\alpha'(m^{(1)})^2=-2|\zeta_2^{(1)}|+\frac{y_1^2}{f_3\mathcal{V}^{1/3}}\approx -\frac{2\alpha'|\kappa_2^{(1)}|}{\pi v_2}+\frac{y_1^2}{f_3\mathcal{V}^{1/3}}\approx  \mathcal{V}^{-1/3}\left(\frac{y_1^2}{f_3}-\frac{2|\kappa_2^{(1)}|}{\pi f_2}\right),\\& \alpha'(m^{(2)})^2=-2|\zeta_3^{(2)}|+\frac{y_2^2}{f_1\mathcal{V}^{1/3}}\approx -2\frac{\alpha'|\kappa_3^{(2)}|}{\pi v_3}+\frac{y_2^2}{f_1\mathcal{V}^{1/3}}\approx  \mathcal{V}^{-1/3}\left(\frac{y_2^2}{f_1}-\frac{2|\kappa_3^{(2)}|}{\pi f_3}\right),
        \label{Eq.tach13}
    \end{split}
\end{equation}
where, $f_i=\alpha'^{-1}v_i/\mathcal{V}^{1/3}$ is a fraction such that $f_1f_2f_3=1$. Here, $y_1$ and $y_2$ are related to the torus complex structure moduli ($U_i)$, which come from the discrete Wilson lines. We observe from eq.(\ref{Eq.tach13}) that the states  ($D7^{(1)}-D7^{(1)}$) and ($D7^{(2)}-D7^{(2)}$) can be elevated to the positive squared mass matter fields by the appropriate choice of $y_1, y_2$ and the values of the magnetic fluxes $\kappa_2^{(1)}$ and $\kappa_3^{(2)}$. In this way, the tachyonic states can be eliminated.\par Next, considering the $D7^{(3)}$-brane stack separations in its transverse direction \cite{Antoniadis:2021lhi,Antoniadis:2022ore}, and taking the example of the doubly-charged state ($D7^{(3)}-D7^{(3)}$), we can get its positive squared mass contribution and thus the total mass become as, 
\begin{equation}
    \begin{split}
       \alpha'(m^{(3)})^2=-2|\zeta_1^{(3)}|+y_3f_3\mathcal{V}^{1/3}\approx -\frac{2\alpha'\kappa_1^{(3)}}{\pi v_1}+{y_3}{f_3\mathcal{V}^{1/3}}\approx  y_3f_3\mathcal{V}^{1/3}-\frac{2|\kappa_1^{(3)}|}{\pi f_1\mathcal{V}^{1/3}} .
    \end{split}
    \label{Eq.tbf}
\end{equation}
In eq.(\ref{Eq.tbf}), the positive and the negative terms correspond to the position of $D7^{(3)}$-brane stack with respect to its orientifold image and the magnetic flux ($\kappa_1^{(3)}$) respectively. For a critical value $\mathcal{V}_c$ of $\mathcal{V}$, it is possible that the overall squared mass $\alpha'(m^{(3)})^2$ becomes negative and thus the mater field is tachyonic. This possibility is interesting so far as cosmological inflation is concerned, as the tachyon will assist to end the inflation and begin the reheating process \cite{Jain:2009ep,Copeland:2002ku, Rashidi:2017chf,Kofman:2001rb,Nautiyal:2018lyq}, in the scenario of hybrid inflation \cite{Antoniadis:2022ore}. We shall write more about that aspect in section 5. \par
The actual masses of the tachyonic fields can be obtained from the F-term and the D-term potentials. This procedure will be highlighted in the next subsections.

\subsection{F-term potential for tachyonic matter fields}
Here, we set up the F-term potential for tachyonic matter fields. In an effective field theory with three D7-brane stacks, there are maximum nine unnormalised untwisted fields ($\Phi_i^{(a)}, i,a=1,2,3$)\footnote{We follow the conventions of ref.\cite{Antoniadis:2022ore} and assume to be  $\Phi_i^{(a)}$ dimensionless}. Here, $a$ designates a particular brane stack and $i$ denotes three directions $viz.$, one transverse direction and two worldvolume directions \cite{Ibanez:2012zz,Font:2004cx,Antoniadis:2022ore,Antoniadis:2021lhi}. In particular, $\Phi_a^{(a)}$ represents a brane modulus, parametrized by the position of the brane stack $a$ in the transverse direction. The remaining two ($\Phi_i^{(a)}, i\neq a$) are the matter fields in the two internal planes of the worldvolume of the $D7^{(a)}$-brane stack. In the last subsection, we have explained how a matter field can become tachyonic. Such a tachyonic field acquires an F-term scalar potential due to the stretching of the strings between the separated   $D7^{(a)}$-brane stack and its orientifold image. In this context, a trilinear superpotential for the tachyonic field, $\Phi_i^{(a)}$, in the $N=1$ supersymmetry, which is a truncation of the $N=4$ supersymmetry, can be written   \cite{Lust:2004fi,Lust:2004dn,Camara:2004jj,Font:2004cx,Ibanez:2012zz,Abe:2021uxb,Antoniadis:2022ore,Lust:2004cx} as
\begin{equation}
    \mathcal{W}_{\Phi_i^{(a)}}=\omega_{ijk}\Phi_i^{(a)}\Phi_j^{(a)}\Phi_k^{(a)}
    \label{Eq.supm}
\end{equation}
 where, $\omega_{ijk}$ is a trilinear coupling. In this case,  $\Phi_j^{(a)}$ and $\Phi_k^{(a)}$ ($j\neq k\neq a)$ are related to the tachyonic scalar and its charge conjugate.   A K\"ahler potential for the matter fields at tree-level can be written \cite{Antoniadis:2021lhi,Lust:2004fi,Font:2004cx,Ibanez:2012zz} in the small magnetic fields limit as,
\begin{equation}
    \begin{split}
       \kappa^2 \mathcal{K}^{(a)}=&-\ln{\left[\frac{1}{(2i)^2}(S-\Bar{S})\left(\mathcal{T}_a-\mathcal{\Bar{T}}_a\right)\left(\frac{1}{2i}\left(S-\Bar{S}\right)\left(U_a-\Bar{U}_a\right)-\left|\Phi_{a}^{(a)}\right|^2\right)\right]}\\&-\ln{\left[\frac{1}{(2i)^2}(S-\Bar{S})(U_j-\Bar{U}_j)\left(\mathcal{T}_k-\mathcal{\Bar{T}}_k\right)-\left|\Phi_{j}^{(a)}\right|^2\right]}\\&-\ln{\left[\frac{1}{(2i)^2}(S-\Bar{S})(U_k-\Bar{U}_k)\left(\mathcal{T}_j-\mathcal{\Bar{T}}_j\right)-\left|\Phi_{k}^{(a)}\right|^2\right]},
        \label{Eq.mk}
    \end{split}
\end{equation} where, in the R.H.S, the first term corresponds to the transverse direction and the second and third terms correspond to the worldvolume directions.
If there is no kinetic mixing, then
the K\"ahler metrics ($ K_{i\Bar{i}}^{(a)}$), $a=1,2,3$ 
 can be obtained \cite{Ibanez:2012zz,Font:2004cx,Antoniadis:2022ore}, from eq.(\ref{Eq.mk}) (using the small field $\Phi_k^{(a)}$ approximation in comparison  to the complex structure moduli and the K\"ahler moduli) as,
\begin{equation}
  \kappa^2K_{i\Bar{i}}^{(1)} = \begin{pmatrix} \frac{2i}{\left(S-\Bar{S}\right)\left(U_1-{\Bar{U}}_1\right)} & 0 & 0 \\
0 & \frac{(2i)^2}{(S-\Bar{S})\left(U_2-\Bar{U}_2\right)\left(\mathcal{T}_3-\mathcal{\Bar{T}}_3\right)} & 0 \\
0 & 0 & \frac{(2i)^2}{(S-\Bar{S})\left(U_3-\Bar{U}_3\right)\left(\mathcal{T}_2-\mathcal{\Bar{T}}_2\right)} \\
\end{pmatrix},
\label{Eq.mt1}
\end{equation}
\begin{equation}
 \kappa^2 K_{i\Bar{i}}^{(2)} = \begin{pmatrix} \frac{(2i)^2}{(S-\Bar{S})\left(U_1-\Bar{U}_1\right)\left(\mathcal{T}_3-\mathcal{\Bar{T}}_3\right)}  & 0 & 0 \\
0 & \frac{2i}{\left(S-\Bar{S}\right)\left(U_2-{\Bar{U}}_2\right)}& 0 \\
0 & 0 & \frac{(2i)^2}{(S-\Bar{S})\left(U_3-\Bar{U}_3\right)\left(\mathcal{T}_1-\mathcal{\Bar{T}}_1\right)} \\
\end{pmatrix},
\label{Eq.mt2}
\end{equation}

\begin{equation}
  \kappa^2K_{i\Bar{i}}^{(3)} = \begin{pmatrix} \frac{(2i)^2}{(S-\Bar{S})\left(U_1-\Bar{U}_1\right)\left(\mathcal{T}_2-\mathcal{\Bar{T}}_2\right)}  & 0 & 0 \\
0 & \frac{(2i)^2}{(S-\Bar{S})\left(U_2-\Bar{U}_2\right)\left(\mathcal{T}_3-\mathcal{\Bar{T}}_3\right)} & 0 \\
0 & 0 & \frac{2i}{\left(S-\Bar{S}\right)\left(U_3-{\Bar{U}}_3\right)}\\
\end{pmatrix}.
\label{Eq.mt3}
\end{equation}
The unnormalized fields ($\Phi_a^{(a)},\Phi_j^{(a)},\Phi_k^{a}$) can be transformed to the canonically normalized fields ($\psi_a^{(a)},\psi_-^{(a)},\psi_+^{(a)}$) through the metrices (eq.(\ref{Eq.mt1}), (\ref{Eq.mt2}) and (\ref{Eq.mt3})), as,
\begin{equation}
    \begin{split}
        &\psi_a^{(a)}=\frac{\Phi_a^{(a)}\kappa^{-1}}{\sqrt{\frac{1}{2i}(S-\Bar{S})(U_a-\Bar{U}_a)}},\\&  \psi_+^{(a)}=\frac{\Phi_j^{(a)}\kappa^{-1}}{\sqrt{\frac{1}{(2i)^2}(S-\Bar{S})(U_j-\Bar{U}_j)(\mathcal{T}_k-\mathcal{\Bar{T}}_k)}},\\& \psi_-^{(a)}=\frac{\Phi_k^{(a)}\kappa^{-1}}{\sqrt{\frac{1}{(2i)^2}(S-\Bar{S})(U_k-\Bar{U}_k)(\mathcal{T}_j-\mathcal{\Bar{T}}_j)}}.
         \end{split}
\end{equation}
Here, $\psi_a^{(a)}$ is a brane modulus and   $\psi_-^{(a)}$ is a tachyonic scalar while its charged conjugate is $\psi_+^{(a)}$. The superpotential eq. (\ref{Eq.supm}) can be expressed in terms of the canonically normalized fields and, thus, we get a simplified form \cite{Antoniadis:2022ore,Antoniadis:2021lhi,Font:2004cx}, 

\begin{equation}
    \mathcal{W}_{\mathrm{tach}}^{(a)}=\mathcal{C}^{(a)}\psi_a^{(a)}\psi_-^{(a)}\psi_+^{(a)}, 
    \label{Eq.msu}
\end{equation}
where, 
\begin{equation}
   \mathcal{C}^{(a)}=\kappa^3\left({\frac{1}{(2i)^5}\left(S-\Bar{S}\right)^3\left({U}_a-{\Bar{U}}_a\right)\prod_{j\neq a}\left(\mathcal{T}_j-\mathcal{\Bar{T}}_j\right)\left({U}_j-{\Bar{U}}_j\right)}\right)^{1/2} 
\end{equation}
  arises through canonical normalization of the fields.
\par
In the large volume limit, with small magnetic fields, the exponential of K\"ahler potential in eq.(\ref{Eq.mk}) becomes,
\begin{equation}
    \begin{split}
        e^{\kappa^2\mathcal{K}^{(a)}}&\approx \frac{1}{(S-\Bar{S})^4\prod_{k=1}^{3}\frac{1}{(2i)^7}(U_k-\Bar{U}_k)(\mathcal{T}_k-\mathcal{\Bar{T}}_k)}(1-\kappa^2\psi_-^{(a)2})^{-1}\times\dots\\& \approx \frac{(1+\kappa^2\psi_-^{(a)2})}{(S-\Bar{S})^4\prod_{k=1}^{3}\frac{1}{(2i)^7}(U_k-\Bar{U}_k)(\mathcal{T}_k-\mathcal{\Bar{T}}_k)}.
        \label{Eq.ep}
    \end{split}
\end{equation}
In eq.(\ref{Eq.ep}), we have considered, for simplicity, only the field $\psi_-^{(a)}$. The total superpotential including the tachyonic effect can be written from eqs.(\ref{eq:non}) and (\ref{Eq.msu}), as
\begin{equation}
    \mathcal{W}_{\mathrm{tot}}=\mathcal{W}_0+\mathcal{W}_{\mathrm{np}}+\mathcal{W}_{\mathrm{tach}}^{(a)}=\mathcal{W}_0 + A_1 e^{ia\mathcal{T}_1} +A_2 e^{ia_2\mathcal{T}_2}+\mathcal{C}^{(a)}\psi_a^{(a)}\psi_-^{(a)}\psi_+^{(a)}.
    \label{Eq.tnon}
\end{equation}
The F-term potential including the tachyonic field can be computed from the supergravity formula in eq.(\ref{Eq:sf}). Using the eqs.(\ref{Eq:sf}), (\ref{Eq.mk}), (\ref{Eq.mt1}), (\ref{Eq.mt2}), (\ref{Eq.mt3}), (\ref{Eq.ep}) and (\ref{Eq.tnon}), we get 
\begin{equation}
    \begin{split}
        V_{{\mathrm{F(tach})}}^{(a)}&\approx\frac{(1+\kappa^2\psi_-^{(a)2})}{(S-\Bar{S})^4\prod_{k=1}^{3}\frac{1}{(2i)^7}(U_k-\Bar{U}_k)(\mathcal{T}_k-\mathcal{\Bar{T}}_k)}\kappa^{-4}\left(\left|\frac{\partial\mathcal{W}_{\mathrm{tot}}}{\kappa\partial\psi_+^{(a)}}\right|^2 +\dots\right)\\&\approx \frac{g_s}{\tau_a}\left|\left<\psi_a^{(a)}\right>\right|^2\left(\left|\psi_-^{(a)}\right|^2+\kappa^2\left|\psi_-^{(a)}\right|^4\right)
        \\&= (m_{\mathrm{F}}^{(a)})^{2}|\psi_-^{(a)}|^2+\lambda_{\mathrm{F}}^{(a)}|\psi_-^{(a)}|^4
        \label{Eq.fm}
    \end{split}
\end{equation}
where, in the second line, we have extracted the part of the potential for the $\psi_-$ only, upto the quartic term. Also, in the second line, $\left<\psi_a^{(a)}\right>$ is the VEV of the D-brane modulus, obtained supersymmetrically. Here,  $(m_{\mathrm{F}}^{(a)})^{2}=\mathcal{Y}_a(U_a)\frac{g_s^2}{\kappa^2\tau_a}$ is the physical mass of the tachyon which comes from the $D7^{(a)}$ brane stack position, $\lambda_{\mathrm{F}^{(a)}}=\kappa^2(m_{\mathrm{F}}^{(a)})^2$ and $\mathcal{Y}_a(U_a)$ is related to the complex D-brane modulus arising due to 
$D7^{(a)}$ brane separation \cite{Antoniadis:2022ore}. 

\subsection{D-term potential with tachyonic matter fields}
In the present geometrical model, the magnetic fields of the brane stacks can give rise to the uplifting D-term potential with non-vanishing matter fields, in the effective field theory. The general formula of the D-term potential in terms of Fayet-Ilioupos parameters  ($\xi^{(a)}$), which is a function of K\"ahler moduli, is given
\cite{Antoniadis:2008uk,Antoniadis:2006eu,Antoniadis:2004pp,Font:2004cx,Ibanez:2012zz, Antoniadis:2022ore,Antoniadis:2006eu} by,
\begin{equation}
    \begin{split}
        V_D^{(a)}&=\sum_{a=1}^3\frac{(g^{(a)})^2}{2}\left(\xi^{(a)}+\sum_n q_n^{(a)}|\phi^{(a)}_n|^2\right)^2\\& \approx \sum_{a=1}^3\frac{1}{2}(g^{(a)})^2(\xi^{(a)})^2-2(g^{(a)})^2\xi^{(a)}|\psi_-^{(a)}|^2+2(g^{(a)})^2|\psi_-^{(a)}|^4 + \dots,
        \label{Eq.md}
    \end{split}
\end{equation}
where, the superscript $n$ stands for the number of charged scalar fields. In eq.(\ref{Eq.md}), we have taken tachyonic contribution with charge $q^{(a)}=-2$ only. Here, $g^{(a)}$ is the gauge coupling which can be written in the limit of large volume  and small magnetic fields \cite{Antoniadis:2022ore,Ibanez:2012zz}, as,
\begin{equation}
    (g^{(a)})^2\approx\frac{g_sv_a}{\mathcal{V}\alpha'}\left|w_j^{(a)}w_k^{(a)}\right|^{-1}\quad\quad a\neq j\neq k \neq a.
    \label{Eq.cp}
\end{equation}
The $(\mathrm{mass})^2$ of the tachyonic field (coefficient of $|\psi_-^{(a)}|^2$) in eq.(\ref{Eq.md}),  which comes from the contribution of the magnetic fields on $D7^{(a)}$-brane stack in eq.(\ref{Eq.tach}), is 
\begin{equation}
    (m_D^{(a)})^2=2(g^{(a)})^2\xi^{(a)}=\frac{2|\zeta_i^{(a)}|}{\alpha'}\approx\frac{2g_s^2\alpha'|\kappa_i^{(a)}|}{\pi\kappa^2\mathcal{V}v_i},
    \label{Eq.dmass}
\end{equation}
 where, $\kappa=g_s\sqrt{\alpha'/\mathcal{V}}$ is the four-dimensional reduced Planck length \cite{Ibanez:2012zz,Antoniadis:2022ore,Font:2004cx}. The Fayet-Iliopoulos parameters ($\xi^{(a)}$) can be obtained from eqs.(\ref{Eq.dmass}) and (\ref{Eq.cp}) as,
\begin{equation}
    \xi^{(a)}\approx \frac{g_sv_j}{\pi \kappa^2\mathcal{V}}|\kappa_k^{(a)}|\left|w_j^{(a)}w_k^{(a)}\right|.
    \label{Eq.x2}
\end{equation}

Now, using the eqs.(\ref{Eq.dmass}), 
 (\ref{Eq.x2}), we obtain the D-tem potential from eq.(\ref{Eq.md}), upto the quartic term of tachyonic field ($\psi_-^{(a)}$) as,
\begin{equation}
    V^{(a)}_{D}\approx\frac{\alpha'^3}{\kappa^4\mathcal{V}}\left(\frac{d_1}{v_1v_2^2}+\frac{d_2}{v_2v_3^2}+\frac{d_3}{v_3v_1^2}\right)-(m_D^{(a)})^2|\psi_-^{(a)}|^2+2(g^{(a)})^2 |\psi_-^{(a)}|^4,
    \label{Eq.d2c}
\end{equation}
where, 
\begin{equation}
    d_a=\frac{g_s^3}{2}|w_j^{(a)}w_k^{(a)}||\frac{\kappa_k^{(a)}}{\pi}|^2, \quad a=1,2,3
    \label{Eq:DAP}
\end{equation}
 are the D-term parameters. Now, the D-term potential (eq.(\ref{Eq.d2c})) can be expressed in terms of four cycles moduli ($\tau_a$) instead of two cycles moduli ($v_a$) through the relation $\tau_a=\mathcal{V}\alpha'/v_a$ \cite{Antoniadis:2018hqy,Leontaris:2022rzj,Antoniadis:2021lhi} and thus we get,
\begin{equation}
    V^{(a)}_{D}=\frac{1}{\kappa^4}\left(\frac{d_1\tau_2}{\mathcal{V}^2\tau_3}+\frac{d_3\tau_1^2\tau_3}{\mathcal{V}^4}\right) +\frac{V'_{\mathrm{up}}}{\kappa^4}-(m_D^{(a)})^2|\psi_-^{(a)}|^2+\lambda^{(a)}_{D} |\psi_-^{(a)}|^4,
    \label{Eq.mdt}
\end{equation}
where, $V'_{\mathrm{up}}=d_2/\tau_1^2\tau_2$ is a constant uplifting term ( $i.e.$, independent of internal volume  $\mathcal{V}$) and $\lambda^{(a)}_{D}=2(g^{(a)})^2$ is the quartic coupling term. Eq.(\ref{Eq.mdt}) can be compared with eq.(\ref{Eq.D}), which was the D-term potential without matter fields.

\section{Inflaton potential with tachyonic matter fields}
The total effective potential can be obtained from eqs.(\ref{Eq.fk}),(\ref{Eq.fm}) and (\ref{Eq.mdt}), as
\begin{equation}
    \begin{split}
        V_\mathrm{eff}(\mathcal{V},\tau_3,\psi_-^{(a)})&=\frac{\left(\epsilon\mathcal{W}_0\right)^2}{\kappa^4}\left(\frac{7\left(\xi+2\eta\ln\mathcal{V}\right)-4\mathcal{V}}{2\mathcal{V}^3}-\frac{17\eta\xi\ln\mathcal{V}}{\mathcal{V}^4}\right)\\&+\frac{1}{\kappa^4}\left(\frac{d_1\tau_2}{\mathcal{V}^2\tau_3}+\frac{d_3\tau_1^2\tau_3}{\mathcal{V}^4}\right) +\frac{V'_{\mathrm{up}}}{\kappa^4}+\frac{1}{2}(m^{(a)})^2|\psi_-^{(a)}|^2+\frac{\lambda^{(a)} }{4}|\psi_-^{(a)}|^4
        \label{Eq.teff}
    \end{split}
\end{equation}
where, $(m^{(a)})^2=2((m_F^{(a)})^2-(m_D^{(a)})^2)$ is the effective mass of the tachyon and $\lambda^{(a)}=4(\lambda_{D}^{(a)}+\lambda_{F}^{(a)})$. Here, $a=1,2,3$ stand for three tachyons corresponding to three D7-brane stacks.
The minimization of the effective potential with respect to the $\tau_3$ modulus gives the value of the $\tau_3$ at the minimum of the potential as $(\tau_3)_{\mathrm{min}}$:
\begin{equation}
    \begin{split}
        \frac{\partial V_{\mathrm{eff}}}{\partial\tau_3}=0 \quad\Longrightarrow\quad \left(\tau_3\right)_{\mathrm{min}}=\left(\frac{d_1}{d_3}\right)^{1/2}\frac{\tau_2^{1/2}}{\tau_1}\mathcal{V}.
        \label{Eq.t3m}
    \end{split}
\end{equation} With eq.(\ref{Eq.t3m}), we get the effective potential at the valley of $(\tau_3)_{\mathrm{min}}$ as
\begin{equation}
    \begin{split}
        V_\mathrm{eff}(\mathcal{V},\psi_-^{(a)]})|_{\left(\tau_3\right)_{\mathrm{min}}}=\frac{\left(\epsilon\mathcal{W}_0\right)^2}{\kappa^4}&\left(\frac{7\left(\xi+2\eta\ln\mathcal{V}\right)-4\mathcal{V}}{2\mathcal{V}^3}-\frac{17\eta\xi\ln\mathcal{V}}{\mathcal{V}^4}\right)+\frac{d'}{\kappa^4\mathcal{V}^3} \\&+\frac{V_{\mathrm{up}}}{\kappa^4}+\frac{1}{2}(m^{(a)})^2|\psi_-^{(a)}|^2+\frac{\lambda^{(a)}}{4} |\psi_-^{(a)}|^4,
    \end{split}
    \label{Eq.tamin}
\end{equation}
 where,
\begin{equation}
    d'=2\tau_1\sqrt{\tau_2}(d_1d_3)^{1/2},
    \label{Eq.d'}
\end{equation} is a constant.
Similarly, the minimization of the effective potential (eq.(\ref{Eq.tamin})) with respect to tachyon ($\psi_-^{(a)}$) gives,
\begin{equation}
   \begin{split}
        \frac{\partial V_{\mathrm{eff}}}{\partial\psi_-^{(a)}}=0 \quad \Longrightarrow \left(\psi_-^{(a)}\right)_{\mathrm{min}}&= \pm \frac{|(m^{(a)})(\psi_-^{(a)})|}{\sqrt{\lambda}}, \quad\text{when}\quad
        (m^{(a)})^2(\psi_-^{(a)})<0,\\&=0, \quad\quad\quad\quad \quad\quad\quad\text{when}
\quad  (m^{(a)})^2(\psi_-^{(a)})>0,
\label{Eq.tachm}
\end{split}
\end{equation}
It is clear that for the second case in eq.(\ref{Eq.tachm}), the tachyonic contribution to the effective potential eq.(\ref{Eq.tamin}) is zero. When the mass of $\psi_-^{(a)}$ becomes tachyonic (first case in eq.(\ref{Eq.tachm})) then we get, 
\begin{equation}
    \begin{split}
        V_\mathrm{eff}(\mathcal{V})|_{\left(\tau_3,\psi_-^{(a)}\right)_{\mathrm{min}}}=\frac{\left(\epsilon\mathcal{W}_0\right)^2}{\kappa^4}&\left(\frac{7\left(\xi+2\eta\ln\mathcal{V}\right)-4\mathcal{V}+q'}{2\mathcal{V}^3}-\frac{17\eta\xi\ln\mathcal{V}}{\mathcal{V}^4}\right) \\&+\frac{V_{\mathrm{up}}'}{\kappa^4}+\frac{(m^{(a)})^4(\psi_-^{(a)})}{4\lambda^{(a)}}(\mathcal{V})
    \end{split}
\end{equation} where, $q'=\frac{2d'}{(\epsilon\mathcal{W}_0)^2}$ is a positive D-term parameter.
\par
(i) \underline{Tachyon due to $D7^{(1)}$-brane stack separation}: First, we consider the $D7^{(1)}$ brane separation and the magnetic field ($H_2^{(1)}$) on $D7^{(1)}$-brane stack is turned on in the second internal plane $T_2^2$. The effective mass of the canonically normalized tachionic field ($\psi_-^{(1)}$) is obtained  from eqs. (\ref{Eq.dmass}), 
 (\ref{Eq.fm}) and (\ref{Eq.t3m}) as,

\begin{equation}
    \begin{split}
&\frac{1}{2}(m^{(1)})^2=\frac{\mathcal{Y}_1(U_1)g_s^2}{\kappa^2 \tau_1}\left(1-\left(\frac{\mathcal{V}_{c_1}}{\mathcal{V}}\right)^2\right),
\label{Eq.m1p}
    \end{split}
\end{equation}
where, 
\begin{equation}
    \mathcal{V}_{c_1}=\left(\frac{2|\kappa_2^{(1)}|\tau_1\tau_2}{\pi\mathcal{Y}_1(U_1)}\right)^{1/2},
    \label{Eq.vc1}
\end{equation}
 is the critical volume. It is to be noted that when $\mathcal{V}_{c_1}>\mathcal{V}$, $(m^{(1)})^2<0$ 
\par
The coefficient of the quartic term of the potential and the value of the potential at the minimum point in the direction of $\psi_-^{(1)}$ are respectively given by
\begin{equation}
    \begin{split}
\frac{1}{4}\lambda^{(1)}=\frac{g_s}{\tau_1}\left(2+g_s\mathcal{Y}_1(U_1)\right)
\label{Eq.q1t}
    \end{split}
\end{equation}
and

\begin{equation}
    V(\mathcal{V})|_{(\psi_-^{(1)})_{\mathrm{min}}}=\frac{R_1}{\kappa^4}\left(1-\left(\frac{\mathcal{V}_{c_1}}{\mathcal{V}}\right)^2\right)^2,
    \label{Eq.Th1}
\end{equation}
where,
\begin{equation}
   R_1=\frac{\mathcal{Y}_1^2(U_1)g_s^3}{4\tau_1(2+g_s\mathcal{Y}_1(U_1))},
   \label{Eq.r1}
\end{equation}
 is a tuning parameter. Here, we consider $w_2^{(1)},$ $ w_3^{(1)}$ is 1.

(ii) \underline{Tachyon due to $D7^{(2)}$-brane stack separation}:
Considering the $D7^{(2)}$ brane separation and the magnetic field ($H_3^{(2)}$)  on the third internal plane $T_3^2$,the quantities describe by  eqs. (\ref{Eq.m1p}),(\ref{Eq.q1t}) and (\ref{Eq.Th1}) are given here as,

\begin{equation}
    \begin{split}
        \frac{1}{2}(m^{(2)})^2&=\frac{g_s^2\mathcal{Y}_2(U_2)}{\kappa^2\tau_2}\left(1-\frac{\mathcal{V}_{c_2}}{\mathcal{V}}\right),
    \end{split}
    \label{Eq.tm2}
\end{equation}
\begin{equation}
    \begin{split}
         \frac{1}{4}\lambda^{(2)}=\frac{g_s}{\tau_2}\left(2+g_s\mathcal{Y}_2(U_2)\right),
    \end{split}
\end{equation}

\begin{equation}
    V(\mathcal{V})|_{(\psi_-^{(2)})_{\mathrm{min}}}=\frac{R_2}{\kappa^4}\left(1-\frac{\mathcal{V}_{c_2}}{\mathcal{V}}\right)^2,
    \label{Eq.Th2}
\end{equation}
where,
\begin{equation}
    \mathcal{V}_{c_2}=\left(\frac{2|\kappa_3^{(2)}|\tau_2}{\pi\tau_1\mathcal{Y}_2(U_2)}\right)\sqrt{\frac{d_1\tau_2}{d_3}}, \quad \mathrm{and},\quad R_2=\frac{\mathcal{Y}_2^2(U_2)g_s^3}{4\tau_1(2+g_s\mathcal{Y}_2(U_2))}.
    \label{Eq.vc2r2}
\end{equation}
 $\mathcal{V}_{c_2}$ is the critical volume of the second tachyon and $R_2$ is a tuning parameter. Here, $w_1^{(2)}=w_3^{(2)}=1$
\par
(iii) \underline{Tachyon due to $D7^{(3)}$-brane stack separation}: In this case, we consider the the magnetic field ($H_1^{(3)}$) on the first internal plane $T_1^2$. The equations for the effective mass, quartic coupling and the minimum value of the potential are,

\begin{equation}
    \begin{split}
        \frac{1}{2}(m^{(3)})^2&=\frac{g_s^2\mathcal{Y}_3(U_3)}{\kappa^2\mathcal{V}}\sqrt{\frac{d_3\tau_1^2}{d_1\tau_2}}\left(1-\frac{\mathcal{V}_{c_3}}{\mathcal{V}}\right),
    \end{split}
    \label{Eq.tm3}
\end{equation}

\begin{equation}
    \begin{split}
         \frac{1}{4}\lambda^{(3)}=\frac{g_s\tau_1}{\mathcal{V}}\sqrt{\frac{d_3}{\tau_2d_1}}\left(2+g_s\mathcal{Y}_3(U_3)\right),
    \end{split}
\end{equation}

\begin{equation}
    V(\mathcal{V})|_{(\psi_-^{(3)})_{\mathrm{min}}}=\frac{R_3}{\kappa^4\mathcal{V}}\left(1-\frac{\mathcal{V}_{c_3}}{\mathcal{V}}\right)^2
    \label{Eq.Th3}
\end{equation}
where,
\begin{equation}
   \mathcal{V}_{c_3}=\frac{2|\kappa_1^{(3)}|}{\pi\mathcal{Y}_3(U_3)}\sqrt{\frac{d_1\tau_2}{d_3}},\quad\mathrm{and} \quad R_3=\frac{\mathcal{Y}_3^2(U_3)g_s^3\tau_1}{4(2+g_s\mathcal{Y}_3(U_3))}\sqrt{\frac{d_3}{d_1\tau_2}}.
   \label{Eq.vc3r3}
\end{equation}
 $\mathcal{V}_{c_3}$ is the critical volume for the  third tachyon and
 $R_3$ is a tuning parameter. Here, $w_2^{(3)}=w_1^{(3)}=1$
\par
Taking three tachyons together, the effective potential of the canonically normalized scalar field ($\it{i.e.},$ inflaton field in eq.(\ref{Eq.canonical})) should be written as,

\begin{equation}
    \begin{split}
        V\left(\phi,\phi_{c_1},\phi_{c_2},\phi_{c_3}\right)=\frac{\left(\epsilon\mathcal{W}_0\right)^2}{2\kappa^4}&e^{-\frac{3}{\sqrt{2}}\phi}\left(q -4e^{\frac{1}{\sqrt{2}}\phi}+\left(7-17\xi e^{-\frac{1}{\sqrt{2}}\phi}\right)\sqrt{2}\eta\phi\right)+\frac{V_{\mathrm{up}}'}{\kappa^4}\\&-\frac{R_1}{\kappa^4}\left(1-e^{\sqrt{2}(\phi_{c_1}-\phi)}\right)^2-\frac{R_2}{\kappa^4}\left(1-e^{\frac{(\phi_{c_2}-\phi)}{\sqrt{2}}}\right)^2\\&-\frac{R_3}{\kappa^4}e^{-\phi/\sqrt{2}}\left(1-e^{\frac{(\phi_{c_3}-\phi)}{\sqrt{2}}}\right)^2,
    \end{split}
    \label{Eq.3tach}
\end{equation}
where, 
\begin{equation}
    \begin{split}
        q=q'+7\xi=\frac{4\tau_1\sqrt{\tau_2}\sqrt{d_1d_3}}{(\epsilon\mathcal{W}_0)^2}+7\xi \quad \mathrm{and}\quad V'_{\mathrm{up}}=\frac{d_2}{\tau_1^2\tau_2}.
    \end{split}
    \label{Eq.qup}
\end{equation}
  Again, $q$ and $V_{\mathrm{up}}'$ are $\mathcal{V}$-independent constants.
 Looking the eqs. (\ref{Eq.pup}) and (\ref{Eq.qup}), we observe that if we assume,
 \begin{equation}
     a_1\approx a_2,
     \label{Eq.apra1a2}
 \end{equation}
  and thus, from eq.(\ref{Eq.a1a2}),
  \begin{equation}
     \tau_1\approx \tau_2 
     \label{Eq.apt1t2}
  \end{equation}
 and interchange the roles of $d_1$ and $d_2$, then we can write,
 \begin{equation}
     q=p \quad\Longrightarrow \quad \frac{2d'}{(\epsilon\mathcal{W}_0)^2}+7\xi =\frac{2d}{(\epsilon\mathcal{W}_0)^2}+7\xi\quad\Longrightarrow d'=d
     \label{qep}
 \end{equation}
 and
 \begin{equation}
     V_{\mathrm{up}}=V'_{\mathrm{up}}.
     \label{VdV}
 \end{equation}
  This will ensure that the inflaton potential in eq. (\ref{Eq.wotach}) is the same as that in eq.(\ref{Eq.3tach}), if we neglect the tachyonic part in the latter. Thus, the stated simple and reasonable assumptions allow us to maintain the inflationary phase ($i.e.$, slow-roll plateau) without tachyons, validating the results of our earlier work \cite{Let:2023dtb}.

\section{Results and discussion}
Now, we discuss the overall consequences of the inclusion of  open-string tachyonic fields on the inflaton potential. It is evident that only the addition of tachyonic matter fields shifts the minima of the inflaton potential from its original position. We are, actually, interested in these cases. \par

Under the condition $\phi<\phi_{c_i}$ (or $\mathcal{V}<\mathcal{V}_{c_i}$), the tachyonic effects set in via a phase transition (see eqs. (\ref{Eq.tachm}), 
 (\ref{Eq.m1p}), (\ref{Eq.tm2}) and (\ref{Eq.tm3})), resulting in the shift of the minimum of inflaton potential. However, in the inflationary phase, where the inflaton field $\phi$ rolls down,  the matter fields are non-tachyonic.  These zero-VEV fields are assumed to be stabilized with large masses and, therefore, they do not influence any part of the inflaton potential. \par
 The inflationary scenario having the inflaton field along with the tachyonic matter fields is termed as  $hybrid$ $inflation$ \cite{Ahmed:2022dhc,Antoniadis:2021lhi,Gong:2022tfu}. In this scenario, The tachyon is treated as a $waterfall$ 
$field$. A notable feature of this model is that the extra field (tachyon or waterfall) is responsible for the end of inflation. Also, the negative contribution of the tachyonic scalar lowers the minimum of the inflaton potential.\par
The coefficients $R_a$ ($a=1,2,3$) of the inflaton potential (eq.(\ref{Eq.3tach})), which are related to the critical value of internal volume, the D-term parameters and the supersymetrically stabilized K\"ahler moduli ($\tau_{1,2}$), can be used as fine-tuning parameters for determining of the minimum value of the inflaton potential and hence for ascertaining the value of the cosmological constant ($\Lambda$) within the framework of the present model. 
From eqs.(\ref{Eq.vc1}), (\ref{Eq.r1}), (\ref{Eq.vc2r2}), (\ref{Eq:DAP}), (\ref{Eq.apt1t2}), (\ref{Eq.d'}) and (\ref{Eq.vc3r3}), we have,
\begin{equation}
    \begin{split}
        R_1=&\beta_1\frac{d_1}{\mathcal{V}_{c_1}^4}\tau_1^3, \quad \mathrm{where},\quad \beta_1=\frac{2}{2+\mathcal{Y}_1(U_1)g_s},
    \end{split}
    \label{Eq:Fintun1}
\end{equation}
\begin{equation}
    \begin{split}
        R_2=\beta_2\frac{d_2}{\mathcal{V}_{c_2}^2}\frac{d_1}{d_3}, \quad \mathrm{where},\quad \beta_2=\frac{2}{2+\mathcal{Y}_2(U_2)g_s},
    \end{split}
    \label{Eq:fintun2}
\end{equation}
\begin{equation}
    \begin{split}
        R_3=\beta_3\frac{d'}{2\mathcal{V}_{c_3}^2}, \quad \mathrm{where},\quad \beta_3=\frac{2}{2+\mathcal{Y}_3(U_3)g_s}.
    \end{split}
    \label{Eq:fintun3}
\end{equation}
We can see that the parameter $\beta_a$ lies between 0 and 1 (since, $\mathcal{Y}_a(U_a)\geq 0$): for $\beta_a=0$, the tachyonic contribution vanishes and for $\beta_a=1$, the tachyonic contribution is maximum.\par 
Let us now graphically study the tachyonic effects near the minimum of the inflaton potential (eq. \ref{Eq.3tach})). In order to have a clear comparison with our earlier work \cite{Let:2023dtb}, we choose the values: $\epsilon\mathcal{W}_0=-0.75$, $\eta=-0.2$, $\xi=100$, $a_1=a_2=0.1$, $d=14.7885$ and $V_{\mathrm{up}}\approx 4.741\times 10^{-6}$. We also note that, with these values, the minimum value of the potential is $V_{min}=4.76\times 10^{-9}$ which occurs at the $\phi=\phi_{min}=7.986$. 
\par
First, considering the effect of the tachyon coming from the $D7^{(1)}$-brane stack only (ignoring the last two terms in eq. (\ref{Eq.3tach})), we observed that the fine-tuning parameter $R_1$ depends crucially on the critical volume $\mathcal{V}_{c_1}$ of the internal space through eq.(\ref{Eq:Fintun1}). However, $\mathcal{V}_{c_1}$ can not be exactly determined within the present model. Hence, we estimate it with the help of eq.(\ref{Eq.canonical}) and by some physical arguments. From eq.(\ref{Eq.canonical}), we find $\mathcal{V}({\phi=\phi_{min}})=283.4$. The tachyonic effect will start just before the minimum value of the potential is reached, $i.e.$, when $\phi$ is slightly greater than $\phi_{min}$: $\phi=8.1$ (say).  At this value of $\phi$, $\mathcal{V}=307.2$. We accept, $\mathcal{V}_{c_1}$=307.2. The tachyonic effect will show up when $\mathcal{V}<\mathcal{V}_{c_1}$. \par

\par
 Now, from eqs. (\ref{qep}), (\ref{VdV}), (\ref{Eq:DAP}), (\ref{Eq.qup}) and (\ref{Eq.apra1a2}), we get,
 
 \begin{equation}
    \begin{split}
        V_{\mathrm{up}}\approx 4.741\times 10^{-6}\approx\frac{d_2}{\tau_1^3}=\frac{g_s^3}{2\tau_1^3\pi^2}|w_1^{(2)}w_3^{(2)}|\left|\frac{n_3^{(2)}}{w_3^{(2)}}\right|^2.
    \end{split}
    \label{Eq:up}
\end{equation}
and
\begin{equation}
    \begin{split}
      14.7885\approx 2\sqrt{d_1d_3}\tau_1^{3/2}\quad \mathrm{and}\quad d_1=\frac{g_s^3}{2\pi^2}|\kappa_2^{(1)}|^2  
    \end{split}
    \label{Eq:dd}
\end{equation}   
 \par
 
Here, we have chosen the flux number $n_2^{(1)}$ to be $1$. For numerical computations, we may choose the wrapping numbers, the flux numbers, the value of $\tau_1$ modulus and the string coupling constant, which satisfy the above all relations, as, 

\begin{equation}
    \begin{split}
        &w_1^{(2)}=19, \quad w_3^{(2)}=3, \quad w_1^{(3)}=1, \quad w_2^{(3)}=7 \\& n_3^{(2)}=5  \quad n_1^{(3)}=1, \quad \tau_1=59\quad g_s=0.4957. 
    \end{split}
    \label{Eq:pa}
\end{equation} 
In eq.(\ref{Eq:pa}) displays one set of parameters. However, this choice is not unique\footnote{Many parameter sets can be chosen, subject to the constraint eqs. (\ref{Eq:up}) and (\ref{Eq:dd}).}. Now from eq.(\ref{Eq.vc1}), we have,
\begin{equation}
    \mathcal{V}_{c_1}^2=(307.2)^2=\frac{2\tau_1^2}{\pi\mathcal{Y}_1(U_1)}.
\end{equation}

\par
\begin{figure}[H]
    \centering
    \includegraphics[width=12cm]{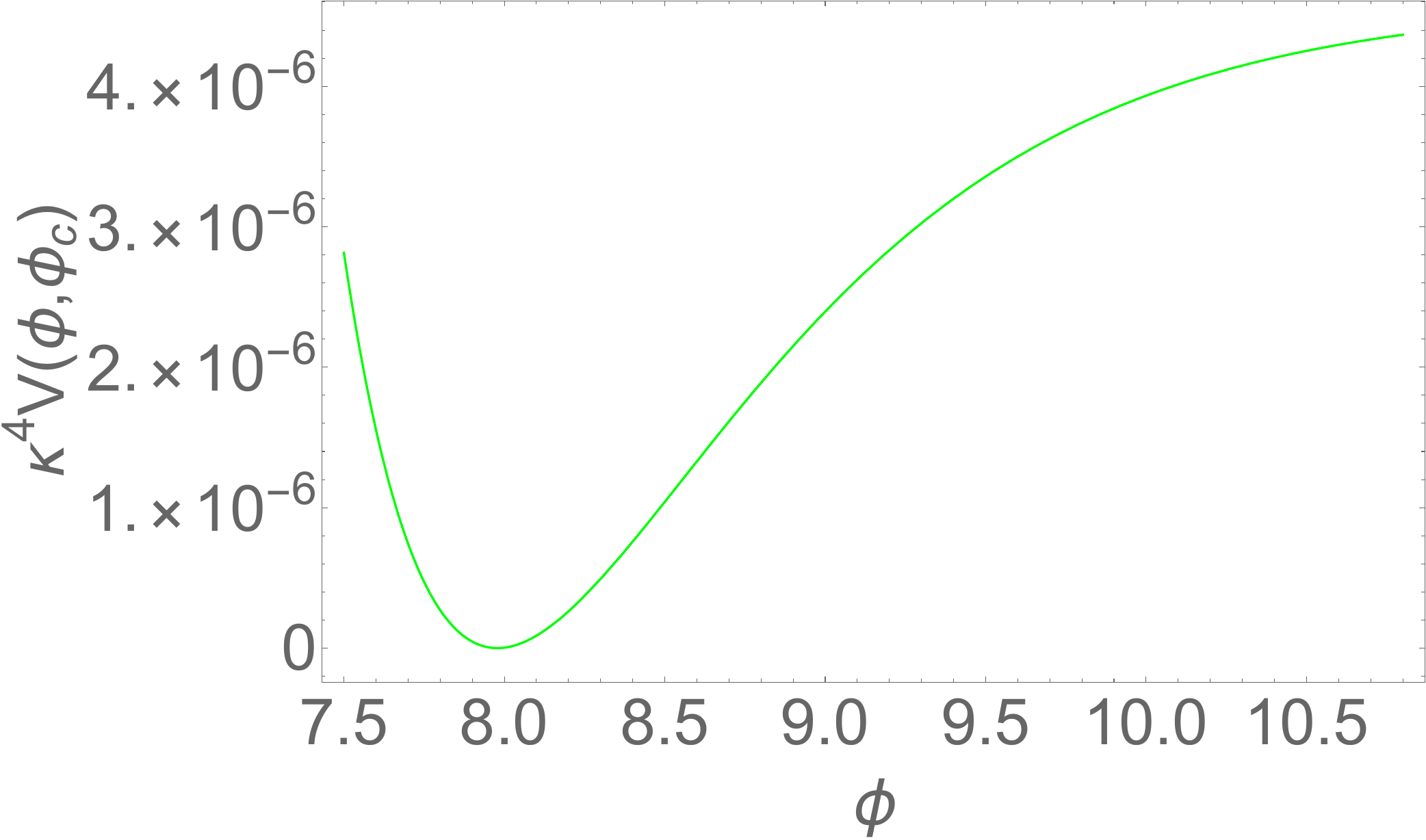}
    
    \caption{Plot of inflaton potentials with the first tachyon only, which comes from $D7^{(1)}$-brane stack separation.}
    \label{fig:1stt1}
\end{figure}

Using the above equations, we calculate $\mathcal{Y}_1(U_1)$ and the fine-tuning parameter $R_1$. Then, in fig.\ref{fig:1stt1}, we have plotted the inflaton potential (eq. (\ref{Eq.3tach})) with one tachyon. Some important parameters for this graph along with $\phi_{min}$ and $V_{min}$ are displayed in Table \ref{tab:1stt}.

\begin{table}[h]
    \centering
    \caption{Useful set of parameters for drawing the inflaton potential (eq.(\ref{Eq.3tach})) with one tachyon coming from $D7^{(1)}$-brane stack.}
    \begin{tabular}{|c|c|c|c|c|c|}
    \hline
          $d_1$& $\mathcal{Y}_1(U_1)$&$\beta_1$& $R_1$&$\phi_{\mathrm{min}}$&$V_{\mathrm{min}}$(in Planck unit)  \\
        \hline
        \vtop{\hbox{\strut $6.17$}\hbox{\strut $\times 10^{-3}$}}&$2.35\times 10^{-2}$&0.9942 &\vtop{\hbox{\strut $1.4146$}\hbox{\strut $\times 10^{-7}$}}&7.9782& $3.274 \times 10^{-11}$\\
        \hline
    \end{tabular}
    
    \label{tab:1stt}
\end{table}

\begin{figure}[H]
    \centering
    
    \includegraphics[width=12cm]{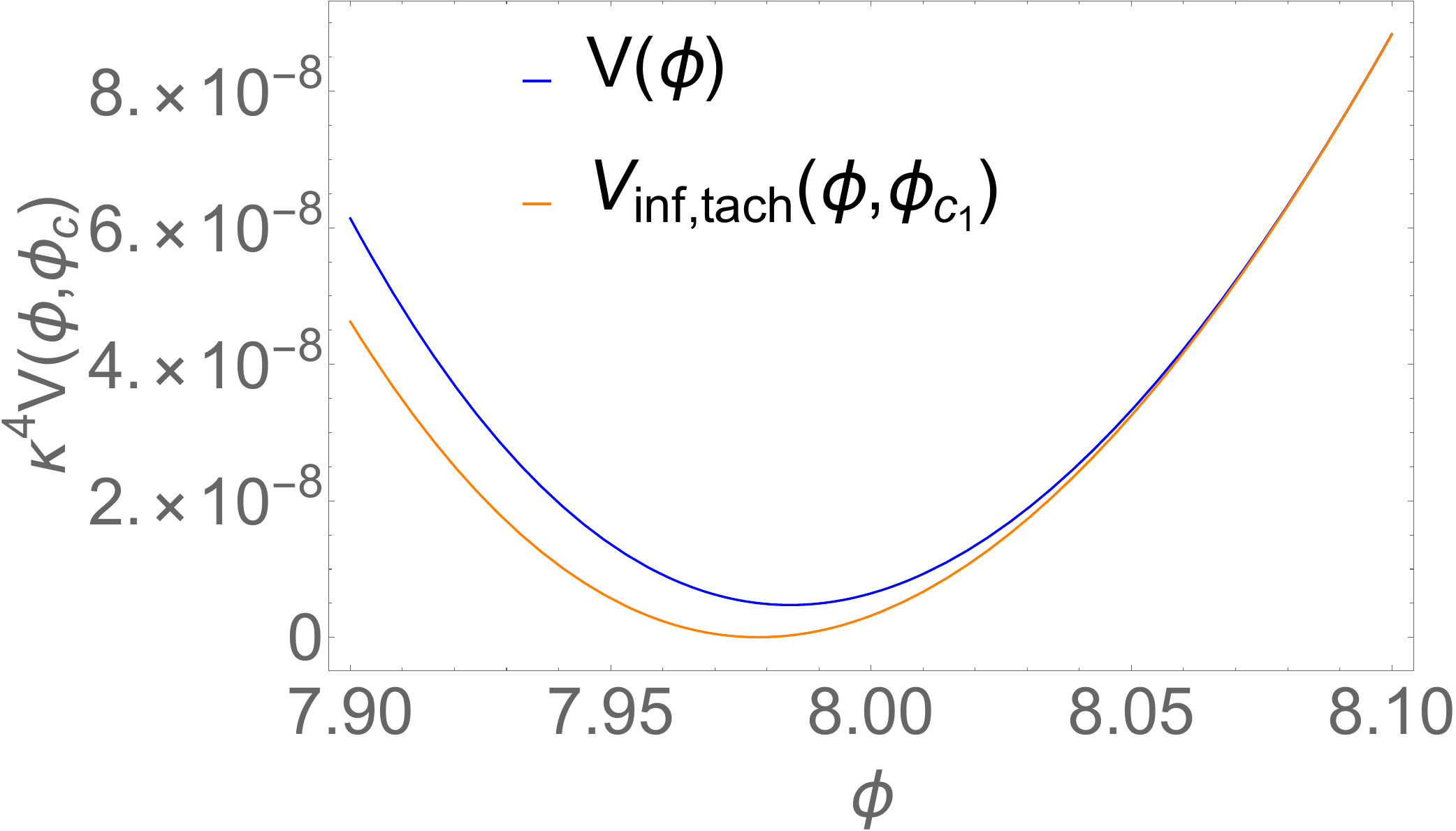}
    \caption{Plot of the bottom portion of the  inflaton potential region  (below the critical value of internal volume): without tachyon $V_{inf}(\phi)$ and with one tachyon $V_{inf,tach}(\phi, \phi_{c_1})$}
    \label{fig:1stt1c}
\end{figure} In fig.\ref{fig:1stt1c}, We have depicted the potential around its minimum at a magnified scale, which can resolve the graph without tachyon from that with one tachyon. It is clear that the net effect of the tachyon is to lower the minimum of the inflaton potential graph. As we have stated earlier $R_1$ can act as a master parameter controlling the value of the minimum. In Table 2, we have shown how $R_1$ can be fine-tuned by the addition of non-zero higher decimal numbers and subject to the accuracy to the calculation tools, one can reach almost zero minimum at the inflaton potential. Since the minimum of the inflaton potential is identified with the cosmological constant in the $dS$ space, the present outcome is consistent with the observed value of the cosmological constant $viz$, $\Lambda\approx 10^{-120} M_P^4$ \cite{Antoniadis:2019doc,SupernovaSearchTeam:1998fmf,SupernovaCosmologyProject:1998vns}. We may also add, here, that only a single tachyon is sufficient in obtaining an almost vanishing potential at the minimum. 
\begin{table}[h]
    \centering
    \caption{Fine-tuning of the parameter $R_1$ (Calculations done by WOLFRAM MATHEMATICA 11.2).}
    \begin{tabular}{|c|c|}
    \hline
        $R_1\times 10^{7}$ & $V_{\mathrm{min}}$ $(M_{\mathrm{P}}^4)$  \\
        \hline
        1.423 &$3.06\times10^{-12}$\\
        \hline
        1.4238 &$2.32\times10^{-13}$\\
        \hline
        1.42386 &$2.04\times10^{-14}$\\
        \hline
        1.423865 &$2.77\times10^{-15}$\\
        \hline
        1.4238657 &$2.92\times10^{-16}$\\
        \hline
        1.423886578 &$9.59\times10^{-18}$\\
        \hline
        1.4238865782 &$2.52\times10^{-18}$\\
        \hline
        1.42388657827 &$4.71\times10^{-20}$\\
        \hline
        1.423886578271 &$1.18\times10^{-20}$\\
        \hline
        1.4238865782713 &$1.18\times10^{-21}$\\
        \hline
        1.42388657827133 &$1.17\times10^{-22}$\\
        \hline
        1.423886578271333 &$1.08\times10^{-23}$\\
        \hline
        1.4238865782713333 &$1.65\times10^{-24}$\\
        \hline
        1.42388657827133333 &0\\
        \hline
    \end{tabular}
    
    \label{tab:vr1}
\end{table}
\par
\begin{figure}[H]
    \centering
    \includegraphics[width=12cm]{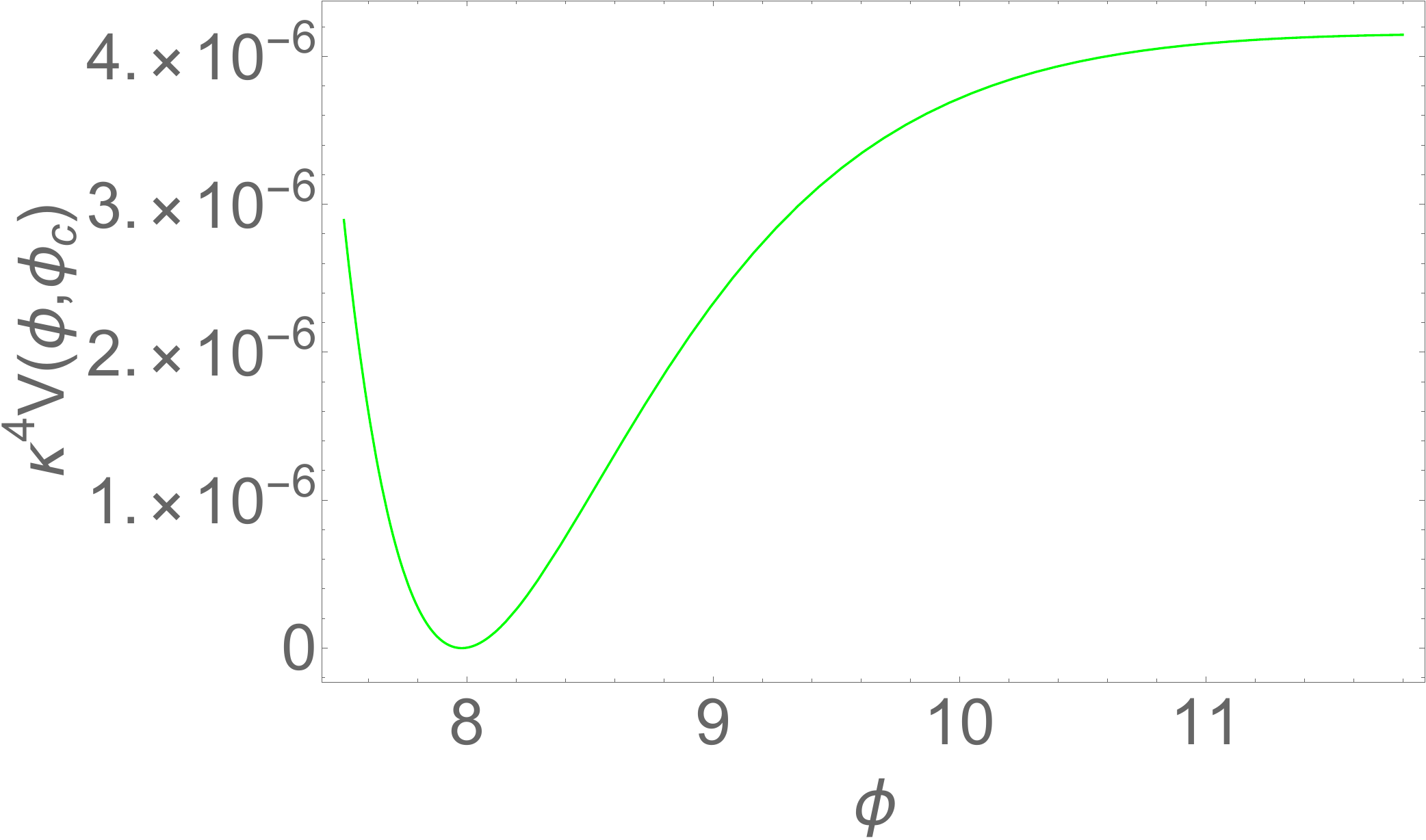}
    
    \caption{Plot of inflaton potentials with one tachyon which comes from $D7^{(2)}$-brane stack separation.}
    \label{fig:2ndd2}
\end{figure}
\begin{figure}[H]
    \centering
    
    \includegraphics[width=12cm]{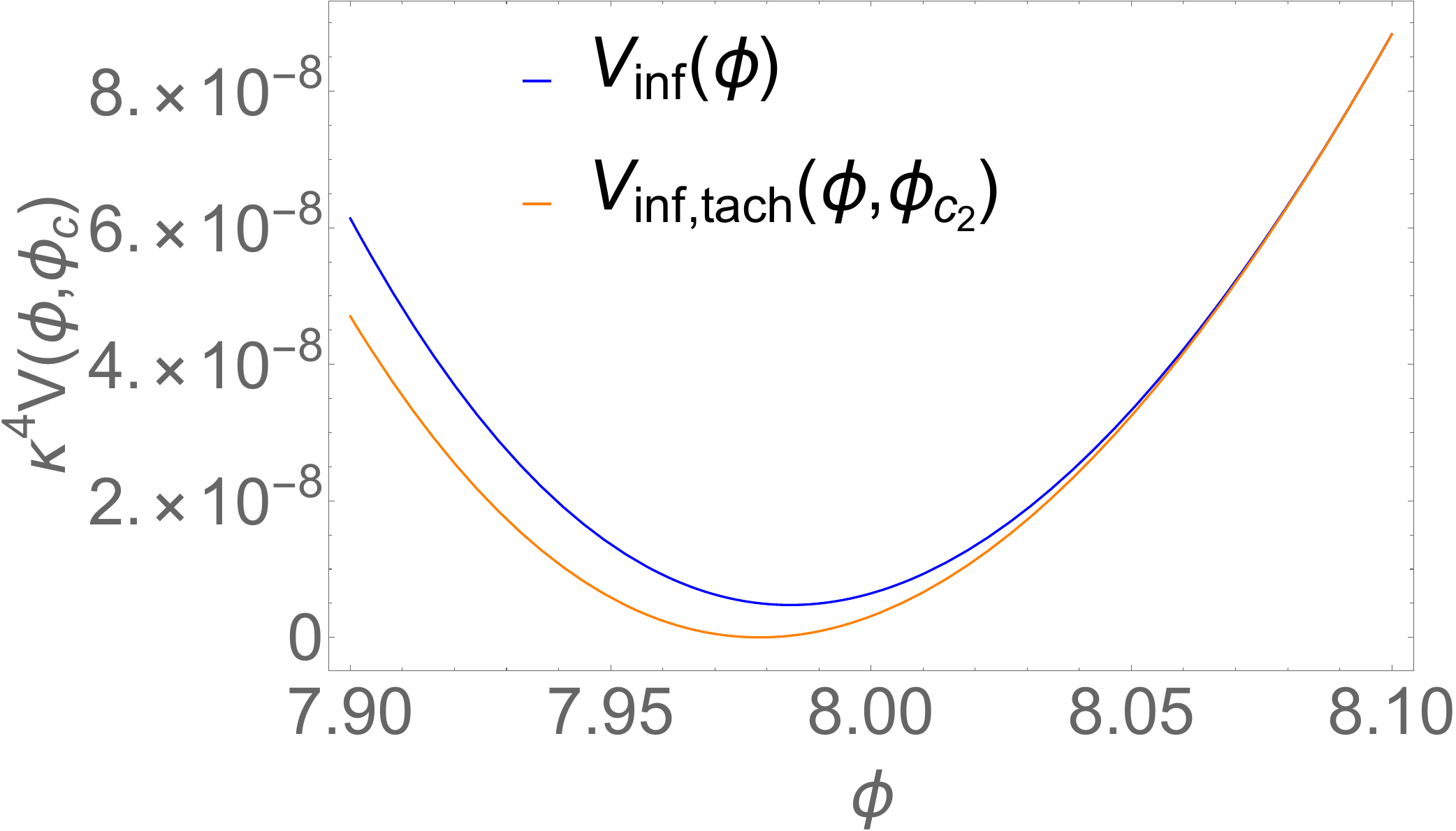}
    \caption{Separation of the minimum (below the $\phi_{c_2}$) of the inflaton potentials with and without a tachyon.}
    \label{fig:2ndd2j}
\end{figure}

Similarly, we examine the case for the tachyon coming from $D7^{(2)}$-brane stack separation only.
 For the sake of comparison with the $D7^{(1)}$-brane stack separation, we choose $\phi_{c_2}=\phi_{c_2}=8.1$ and $\mathcal{V}_{c_2} =\mathcal{V}_{c_1}=307.2$. In this case, we get from eqs. (\ref{Eq:DAP}), (\ref{Eq:up}) and (\ref{Eq.vc2r2}),
\begin{equation}
    \begin{split}
        d_2=4.741\times 10^{-6}\tau_1^3=\frac{g_s^3}{2\pi^2}\quad\mathrm{and}\quad \mathcal{V}_{c_2}=307.2=\frac{2}{\pi\mathcal{Y}_2(U_2)}\sqrt{\frac{d_1\tau_1}{d_3}},
    \end{split}
    \label{Eq:d2vc2}
\end{equation}
where, the flux number $n_3^{(2)}$ has been taken as  $1$. The values of the remaining wrapping numbers, the flux numbers and the $\tau_1$ modulus are chosen as in eq. (\ref{Eq:wftg}) so that they obey the relations in eq. (\ref{Eq:d2vc2}), 

\begin{equation}
    \begin{split}
        &w_1^{(3)}=w_2^{(1)}=1, \quad w_3^{(1)}=107, \quad w_2^{(3)}=29,  \\& n_2^{(1)}= n_1^{(3)}=1, \quad \tau_1=14.995\quad g_s=0.681, 
    \end{split}
    \label{Eq:wftg}
\end{equation}
where, the choice of the parameters is also not unique.  Using eqs. (\ref{Eq:fintun2}), (\ref{Eq:d2vc2}) and (\ref{Eq:wftg}), we can obtain the fine-tuning parameter, $R_2$. In figs. \ref{fig:2ndd2} and \ref{fig:2ndd2j}, we have plotted, respectively, the full inflaton potential with one tachyon and the magnified portion near the minimum, where we compare the potentials without tachyon with that one tachyon. In Table 3, we have listed the important parameters, necessary to plot the potentials.\par
 In Table 4, we display the process of fine-tuning in the parameter $R_2$ which can reduce the value of the minimum and make it almost zero at a certain level, thereby giving a very small value of $\Lambda$.
\begin{table}[h]
    \centering
    \caption{Useful set of parameters for drawing the inflaton potential (eq.(\ref{Eq.3tach})) with one tachyon coming from $D7^{(2)}$-brane stack.}
    \begin{tabular}{|c|c|c|c|c|c|c|c|}
    \hline
          $d_1$&$d_2$&$d_3$& $\mathcal{Y}_2(U_2)$&$\beta_2$& $R_2$&$\phi_{\mathrm{min}}$&$V_{\mathrm{min}}$(in Planck unit)  \\
        \hline
        0.2443&\vtop{\hbox{\strut $1.597$}\hbox{\strut $\times 10^{-2}$}}&0.0662&$1.54\times 10^{-2}$&0.9948 &\vtop{\hbox{\strut $6.2163$}\hbox{\strut $\times 10^{-7}$}}&7.9782& $1.59\times 10^{-12}$\\
        \hline
    \end{tabular}
    
    \label{tab:2ndt}
\end{table}

\par
  
\begin{table}[h]
    \centering
    \caption{Fine-tuning of the parameter $R_2$ (Calculations done by WOLFRAM MATHEMATICA 11.2).}
    \begin{tabular}{|c|c|}
    \hline
        $R_2\times 10^{7}$ & $V_{\mathrm{min}}$ $(M_{\mathrm{P}}^4)$  \\
        \hline
        6.218 &$2.22\times10^{-13}$\\
        \hline
        6.2182 &$6.09\times10^{-14}$\\
        \hline
        6.21827 &$4.58\times10^{-15}$\\
        \hline
        6.218275 &$5.57\times10^{-16}$\\
        \hline
         6.2182756&$7.40\times10^{-17}$\\
        \hline
        6.21827569 &$1.57\times10^{-18}$\\
        \hline
        6.218275691 &$7.69\times10^{-19}$\\
        \hline
        6.2182756919 &$4.47\times10^{-20}$\\
        \hline
        6.21827569195 &$4.49\times10^{-21}$\\
        \hline
        6.218275691955  &$4.65\times10^{-22}$\\
        \hline
        6.2182756919555  &$6.20\times10^{-23}$\\
        \hline
        6.21827569195557 &$5.79\times10^{-24}$\\
        \hline
        6.218275691955577 &$1.65\times10^{-24}$\\
        \hline
        6.2182756919555779 &0\\
        \hline
    \end{tabular}
    \label{tab:my_label}
\end{table}
\par
\begin{figure}[H]
    \centering
    \includegraphics[width=12cm]{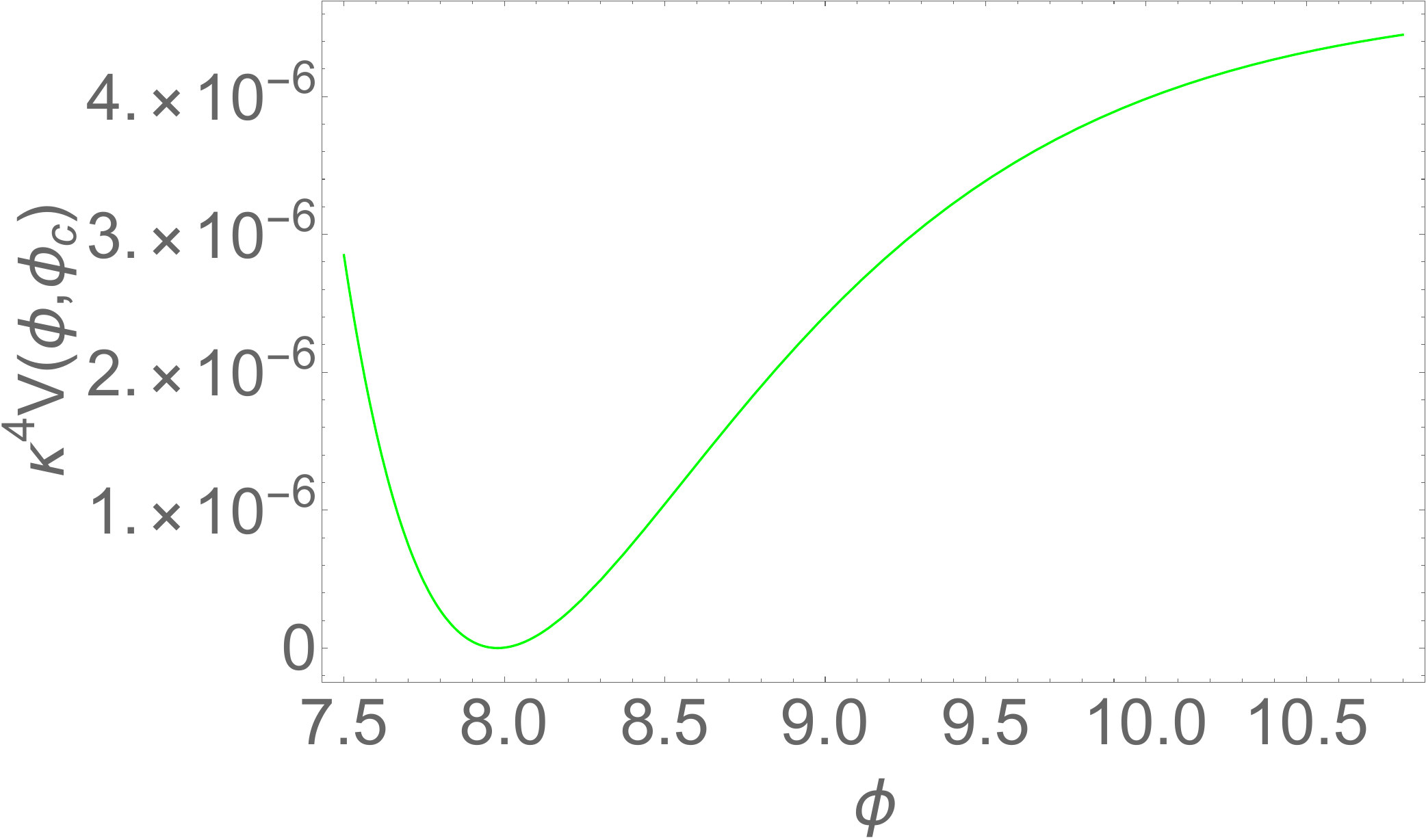}
    
    \caption{Plot of inflaton potentials with the third tachyon only, which comes from $D7^{(3)}$-brane stack separation.}
    \label{fig:3rdd3}
\end{figure}
\begin{figure}[H]
    \centering
    
    \includegraphics[width=12cm]{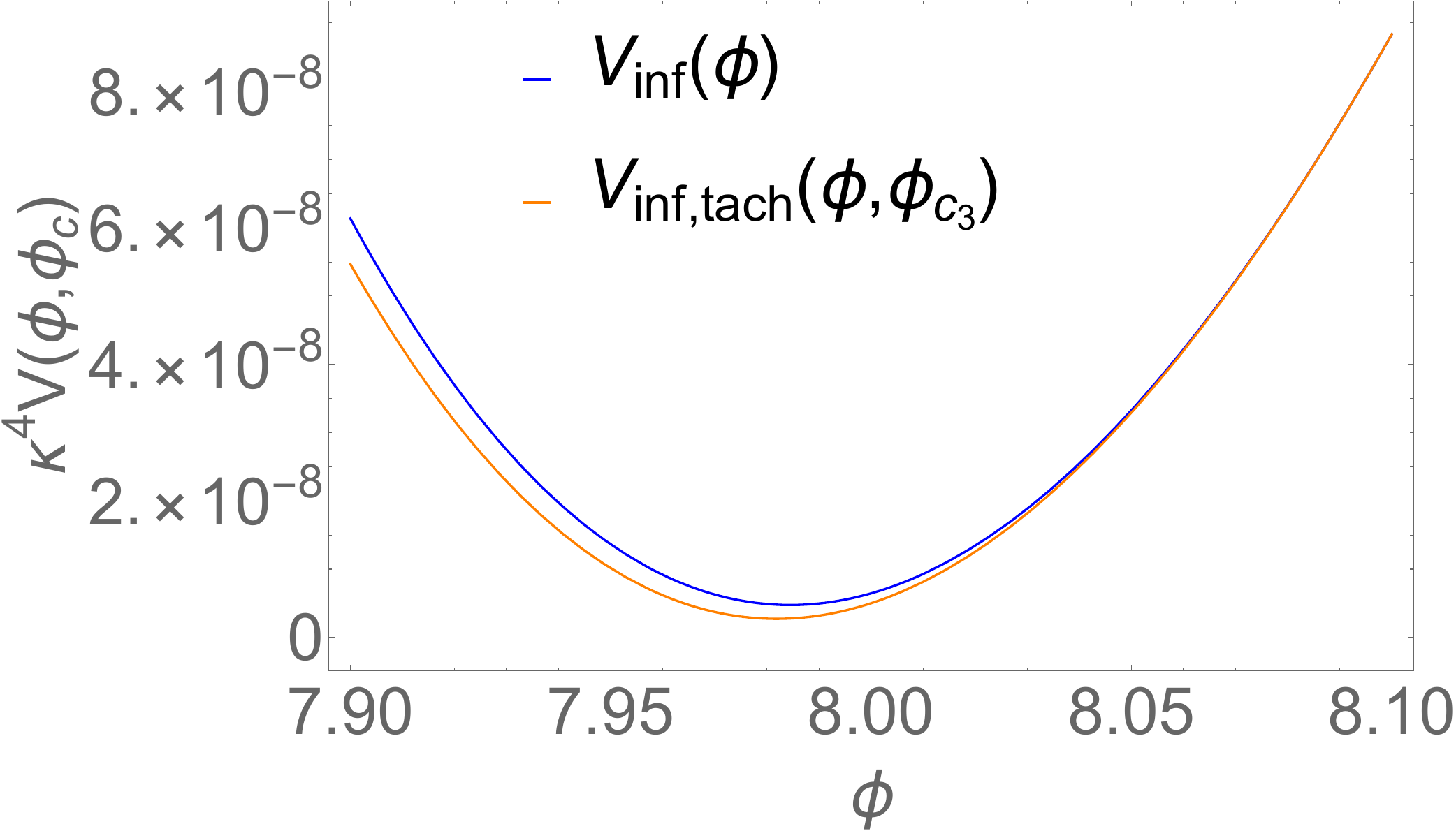}
    \caption{Comparison of the inflaton potentials below the critical value of internal volume: without tachyon and with the third tachyon.}
    \label{fig:3rdd3j}
\end{figure}

 While considering the tachyon coming from the $D7^{(3)}$-brane stack separation only, we must note the difference of this case from the earlier two cases. Here, modulus $\tau_3$ is subjected to only perturbative corrections and not the non-perturbative one. Moreover, it is not supersymmetrically stabilized and acts as the inflaton field \cite{Let:2023dtb}. So far as the parameter $R_3$ is concerned, it can be seen from eq.(\ref{Eq:fintun3}) that once $\mathcal{V}_{c_3}$ is fixed, the former will also be fixed. Then, it can not be treated as a fine-tuning parameter to get at the almost zero value of the cosmological constant $\Lambda$. Nevertheless, in order to draw the inflaton potential graph, we proceed with the following set of parameters,   keeping the critical value of the volume as in the earlier cases:     
\begin{equation}
    \begin{split}
        &w_1^{(3)}=w_2^{(3)}=1,\quad w_1^{(2)}=8, \quad w_3^{(2)}=9, \quad w_3^{(1)}=25, \quad w_2^{(1)}=1,  \\& n_2^{(1)}= n_1^{(3)}=1,\quad n_3^{(2)}=4, \quad \tau_1=30\quad g_s=0.562,\quad d_2=0.128. 
    \end{split}
    \label{Eq:3rdpset}
\end{equation}
As mentioned before, the set of parameters is not unique. Using the eqs. (\ref{Eq:fintun3}), (\ref{Eq:3rdpset}),, (\ref{Eq:DAP}) and (\ref{Eq.Th3}), we obtain the value of  $R_3$ parameter and the minimum value of the potential, as shown in Table \ref{tab:3rdt}.
\begin{table}[h]
    \centering
    \caption{Useful set of parameters for drawing the inflaton potential (eq.(\ref{Eq.3tach})) with one tachyon coming from $D7^{(3)}$-brane stack.}
    \begin{tabular}{|c|c|c|c|c|c|c|}
    \hline
          $d_1$&$d_3$& $\mathcal{Y}_3(U_3)$&$\beta_3$& $R_3$&$\phi_{\mathrm{min}}$&$V_{\mathrm{min}}$(in Planck unit)  \\
        \hline
         0.225&0.009&$1.54\times 10^{-2}$&0.9948 &\vtop{\hbox{\strut $7.7122$}\hbox{\strut $\times 10^{-5}$}}&7.9789& $2.78\times 10^{-9}$\\
        \hline
    \end{tabular}
    
    \label{tab:3rdt}
\end{table}
In fig.\ref{fig:3rdd3}, we have plotted the inflaton potential with one tachyon coming from $D7^{(3)}$-brane stack. In fig.\ref{fig:3rdd3j}, we have depicted the potential around its minimum at a magnified scale, which can resolve the graph without tachyon from that with one tachyon.

\section{Summary and conclusion}
We summarise the main contents of the present paper in the following way: 
\par
(1) We have studied the roles of open-string charged tachyonic scalars in the process of K\"ahler moduli stabilizations over and above the quantum non-perturbative as well as the perturbative corrections within the geometry of the three intersecting magnetized $D7$-brane stacks in type-IIB/F theory. One of the K\"ahler moduli (here, $\tau_3$), which receives only perturbative corrections, is identified as the inflaton field.
\par
(2) The open-string charged states attached to the $D7$-brane stacks receive the non-supersymmetric negative squared masses due to the magnetic fields on the branes. They get physical masses either from the Wilson lines, which can act in the worldvolume direction of the branes, or from the brane separations in the transverse direction. When the brane separations and the magnetic fields are considered together, the untwisted charged states ($D7^{(a)}-D7^{(a)}$) become tachyonic scalars, in case the volume of the internal space is smaller than a critical value.    
 \par
 (3) In the context of effective field theory, the masses of the canonically normalized tachyonic scalars can be computed through the effective potential (sum of the F- and the D-term potentials). The D-term potential comes from magnetic fluxes, whereas the stretching of the string due to brane separation gives rise to the  F-term potential. 
\par
(4) After minimizing the total effective potential of the K\"ahler moduli and the tachyonic scalars with respect to the $\tau_3$ modulus and the tachyonic scalars, we find that a negative contribution for the tachyon adds to the inflaton potential. A tachyonic scalar acts as a waterfall field, in the sense that it lowers the value of the inflaton potential through its negative contribution and thus it helps to end the process of inflation.  \par
(5) In the presence of a tachyonic (waterfall) field at the bottom of the potential, the single-field inflation, which has been described through the moduli stabilization mechanism, enters into the realm of hybrid inflation.    \par
(6) We have explored the cases separately for three tachyons coming from three brane stacks along with the inflaton field. It is observed that tachyons $\psi_-^{(1)}$ and $\psi_-^{(2)}$, coming from the $D7^{(1)}$- and $D7^{(2)}$-brane stacks respectively,  are good candidates to get almost zero value of the potential at the minimum, thereby conforming to the $small$ experimental value of the cosmological constant $\Lambda$.  \par
(7) In this paper, we have considered the simple cases of one tachyon at a time and observed that this does not alter the feature of inflationary plateau, obtained in our previous work \cite{Let:2023dtb}. 
\par
In conclusion, we have shown that the inclusion of tachyonic matter fields, in addition to the perturbative and the non-perturbative corrections to the K\"ahler moduli, in a model of three intersecting D7-brane stacks, is effective in lowering the minimum of the inflaton potential and thus obtaining a very small value of the cosmological constant. It may be noted that the non-perturbative superpotential effects play their role in obtaining the slow-roll plateau of the potential and the simultaneous existence of the tachyon is efficacious in ending the inflation faster and in bringing down the minimum to a lower positive value.

\acknowledgments
The authors acknowledge the University Grants Commission, The Government of India, for the CAS-II program in the Department of Physics, The University of Burdwan. AL acknowledges the CSIR, The Government of India, for granting him the NET (SRF) fellowship.




\bibliographystyle{JHEP}
\bibliography{biblio}
\end{document}